\documentclass[aps,prl,reprint,groupedaddress,showpacs,amsmath,amssymb,twocolumn,floatfix]{revtex4-1}
\usepackage{graphicx}
\usepackage{bm}
\usepackage{color}
\usepackage[normalem]{ulem}
\usepackage{amsmath}
\usepackage{tabularx}

\usepackage[colorlinks=true]{hyperref}  
\hypersetup{
    unicode=false,          % non-Latin characters 
    pdftoolbar=true,        % show Acrobat
    pdfmenubar=true,        % show Acrobat 
    pdffitwindow=false,     % window fit to page when opened
    pdfstartview={FitH},    % fits the width of the page to the window
    pdftitle={Momentum-polarization in BLG},    % title
    pdfauthor={},     % author
    pdfsubject={},   % subject of the document
    pdfcreator={},   % creator of the document
    pdfproducer={}, % producer of the document
    pdfkeywords={} {} {}, % list of keywords
    pdfnewwindow=true,      % links in new window
    colorlinks=true,       % false: boxed links; true: colored links
    linkcolor=magenta, %red,          % color of internal links (change box color with linkbordercolor)
    citecolor=blue,        % color of links to bibliography
    filecolor=magenta,      % color of file links
    urlcolor=blue           % color of external links
} 
\usepackage{cleveref}

\newcommand{\Rxx}{$R_{xx}$}

\newcommand{\Bperp}{$B_{\perp}$}
\newcommand{\Bpara}{$B_{\parallel}$}

\newcommand{\Vpo}{$V^{2\omega}_{\parallel}$}

\usepackage{bbm}

\renewcommand{\vec}[1]{\boldsymbol{#1}}
\newcommand{\Vt}{$V^{2\omega}$}
\newcommand{\arnt}{ARNTM}

\begin{document}

\title{Spontaneous momentum polarization and diodicity in Bernal bilayer graphene}

\author{Jiang-Xiazi Lin$^{1}$}
\author{Yibang Wang$^{1}$}
\author{Naiyuan J. Zhang$^{1}$}
\author{Kenji Watanabe$^{2}$}
\author{Takashi Taniguchi$^{3}$}
\author{Liang Fu$^{4}$}
\author{J.I.A. Li$^{1}$}
\email{jia\_li@brown.edu}

\affiliation{$^{1}$Department of Physics, Brown University, Providence, RI 02912, USA}
\affiliation{$^{2}$Research Center for Functional Materials, National Institute for Materials Science, 1-1 Namiki, Tsukuba 305-0044, Japan}
\affiliation{$^{3}$International Center for Materials Nanoarchitectonics,
National Institute for Materials Science,  1-1 Namiki, Tsukuba 305-0044, Japan}
\affiliation{$^{4}$Department of Physics, Massachusetts Institute of Technology, Cambridge, MA 02139, USA}

\date{\today}

\maketitle

\textbf{The low-temperature phase diagram of multilayer graphene heterostructures is largely defined by the exchange-driven instability that lifts the four-fold isospin degeneracy. Such instability gives rise to the quarter- and half-metal phases, which are key to our understanding of other emergent phenomena. Recent theoretical works shed light on a new type of Coulomb-driven instability. It is proposed that the exchange interaction between trigonal-warping-induced Fermi pockets could induce charge carriers to condense into one of the Fermi pockets, giving rise to a net polarization in the momentum space. Here, we report the observation of spontaneous momentum polarization in Bernal bilayer graphene using angle-resolved nonlinear transport measurement at the second-harmonic frequency. With excellent angular precision, we show that the polar axis of the momentum
polarization is tunable with varying carrier density, electric field, and magnetic field.  The dominating influence of the momentum-space instability reveals a natural connection between broken symmetries, and the isospin degeneracy lifting in the half- and quarter-metal phases. 
}

\begin{figure*}
\includegraphics[width=0.9\linewidth]{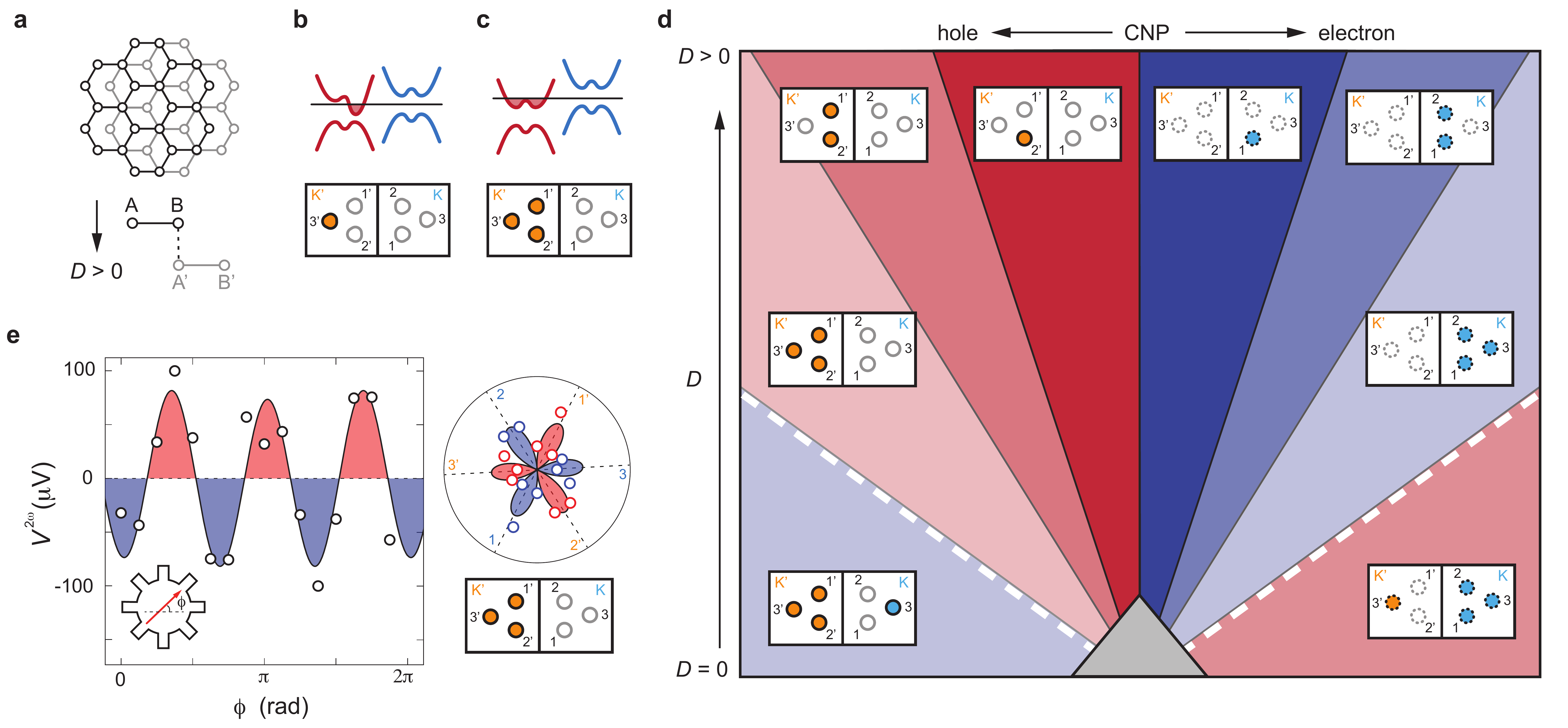}
\caption{\label{fig1}{\bf{Schematic diagram of the momentum-space instability.}} (a) Schematic of Bernal-stacked BLG. Black vertical arrow marks the direction of the positive displacement field $D$.  (b-c) Schematic diagram showing the energy band structure and Fermi surface occupation of (b) a momentum-polarized state where charge carriers condense into one of the trigonally warped pockets, and (c) a valley-polarized state that preserves three-fold rotational symmetry $C_3$.  (d) Schematic diagram of the $n-D$ phase space of BLG. A cascade of momentum-polarized states emerges as charge carriers sequentially occupy separate Fermi pockets.  Areas of the $n-D$ map are marked with different colors based on the underlying pocket occupation.   The white dashed lines indicate the transition boundary between different valley isospin orders.  (e) Nonlinear transport response measured at the second-harmonic frequency, \Vpo, as a function of current flow direction $\phi$.  The measurement is performed at $n=-0.03 \times 10^{12}$ cm$^{-2}$ and $D = 150$ mV/nm. The black solid line is the best fit to the data using Eq.~1, with $V_3 \gg V_1$. Right panel shows a Polar-coordinated plot of the same data. Dashed lines mark azimuth angles corresponding to maximum positive and negative nonlinear responses. Left inset shows the schematic diagram of the sample geometry. Bottom right: schematic diagram of the charge carrier occupation. The three-fold symmetric nonlinear response indicates that carriers occupy all three pockets in valley K', which corresponds to a valley-polarized, yet momentum-unpolarized phase.
 }
\end{figure*}

\begin{figure*}
\includegraphics[width=0.85\linewidth]{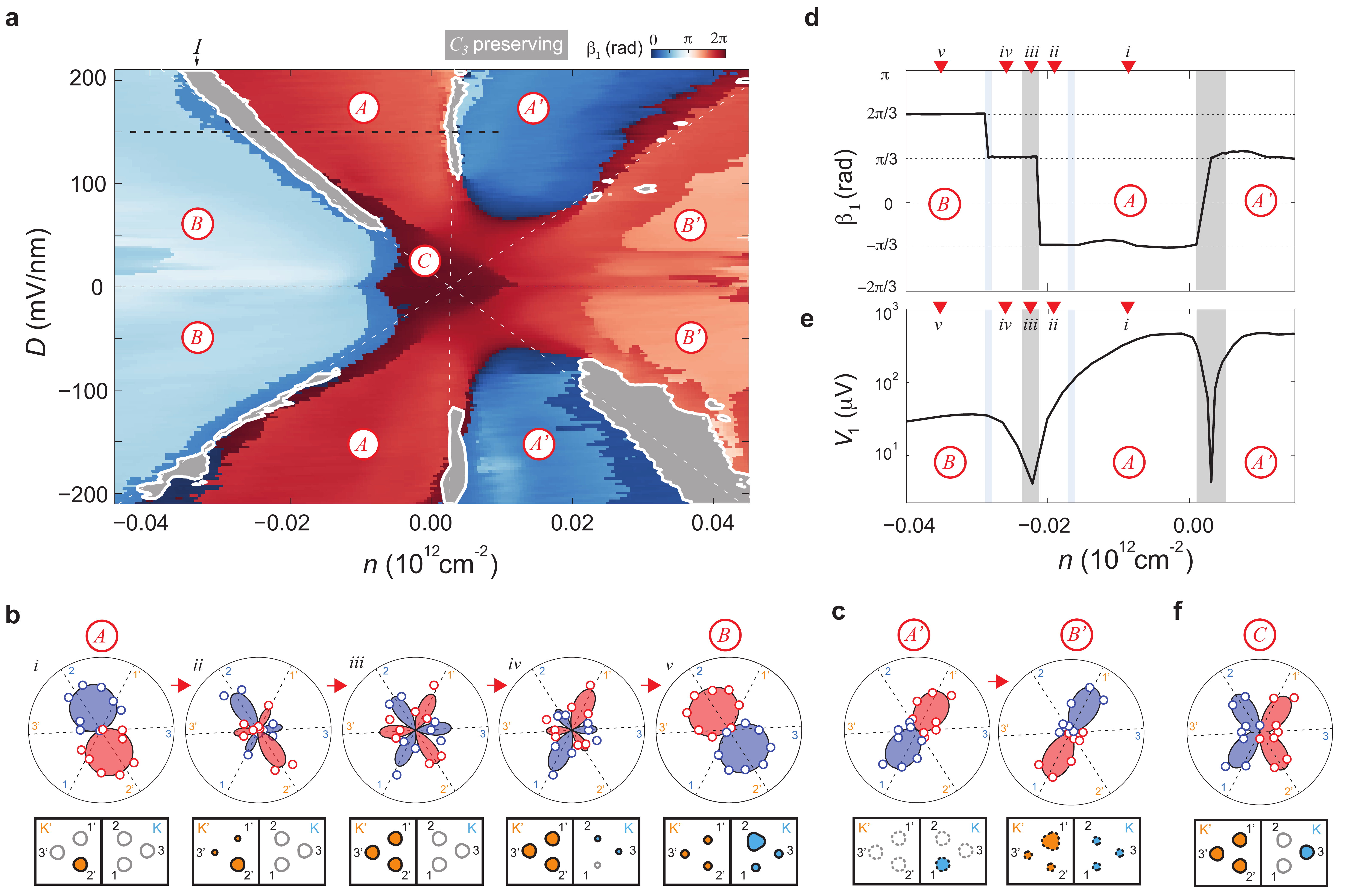}
\caption{\label{fig2}{\bf{Cascade of momentum-polarized states.} }  (a) Polar axis orientation $\beta_1$ as a function of carrier density $n$ and electric field $D$. Gray color indicates  regimes with a predominantly three-fold symmetric nonlinear response. Based on the polar axis orientation, the $n-D$ map is divided into distinct areas with the one-fold symmetric nonlinear response,  which are marked with $A$, $B$, $A'$, $B'$, and $C$. Regimes $A$ and $B$ are separated by the transition boundary \textit{I}. (b-c) Polar-coordinate plots showing the angular dependence of the nonlinear response across the transition boundary between (b) regimes $A$ and $B$, (c) $A'$ and $B'$. The bottom panels show schematic diagrams of possible carrier occupation across different Fermi pockets. Filled (empty) circles denote occupied (unoccupied) Fermi pockets. All measurements in (b-c) are performed at a constant electric field of $D = 150$ mV/nm. (d) Polar-axis orientation $\beta_1$ and (e) $V_1$ as a function of charge carrier density $n$ measured at $D = 150$ mV/nm. The line trace in (d-e) is taken along the black dashed line in Fig.~\ref{fig2}a. (f) The polar-coordinate plot shows the angular dependence of the nonlinear response in regime $C$. The bottom panels show the schematic diagram of a possible carrier occupation.
 }
\end{figure*}

Bernal bilayer graphene (BLG) consists of two graphene layers stacked in the AB form, where the A-sublattice of one layer lies directly over the B-sublattice of the other (Fig.~\ref{fig1}a). Despite the simple lattice structure, BLG hosts an intricate landscape of emergent phenomena. For instance, the evolution of transport properties with varying magnetic field provided a glimpse into the reduced rotational symmetry and broken time-reversal symmetry at zero magnetic field~\cite{Weitz2010BLG,Mayorov2011BLG,Maher2013BLG,Lee2014BLG,Kou2014BLG,Maher2014BLG,Velasco2012BLG}. Furthermore, the application of an electric field flattens the energy band structure near the charge neutrality point (CNP) ~\cite{Zhang2009}. As Coulomb interaction is enhanced under the flat band condition, an exchange-driven instability lifts the isospin degeneracy, giving rise to a cascade of half- and quarter-metal phases ~\cite{Zhou2022BLG,Zhang2022BLG,De2022BLG,Seiler2022BLG}. The isospin-degeneracy lifting in Bernal bilayer resembles the behaviors observed in rhombohedral trilayer graphene ~\cite{Chen2020ABC,Zhou2021ABC,Zhou2021ABCSC}. Since their discovery, many efforts have been devoted to unraveling the nature of the isospin-degeneracy lifting and its connection with the superconducting phases ~\cite{Szabo2022BLG,Chou2022acoustic,Ghazaryan2021ABC,Lu2022ABC,Dai2021ABC}.  
Recently, theoretical works shed light on a new type of Coulomb-driven instability in the quarter and half-metal states of moir\'eless bilayer and trilayer graphene.  Based on a three-pocket model, it is argued that the exchange interaction could induce a spontaneous condensation of charge carriers in the momentum space ~\cite{Dong2021momentum,Huang2022momentum,Jung2015momentum}. This phenomenon results from the process where carriers flock to occupy one of the trigonal-warping-induced pockets (Fig.~\ref{fig1}b). 

%previous Such instability provides a novel interplay between spontaneous symmetry breaking and isospin-ordered phases. Together, the exchange-driven instability in the momentum space offers an important extra dimension in the electronic orders of BLG.

According to the theoretical model, spontaneous momentum polarization in BLG is described by a characteristic phase diagram, as shown  in Fig.~\ref{fig1}d ~\cite{Dong2021momentum}. Different colors in the schematic diagram mark regimes where $1$, $2$, $3$, and $4$ Fermi pockets are sequentially occupied as charge carriers are added to the system.  As an applied electric field flattens the energy band, the density of state in each pocket is enhanced and the density range between adjacent transitions increases. The resulting fan-like phase diagram is an important signature for identifying the Coulomb-driven instability in the momentum space. According to the scheme of sequential occupation, all three pockets in one valley must be occupied before charge carriers start to populate the opposite valley. 
As such, the cascade of momentum-polarized states is naturally intertwined with the sequence of isospin-ordered phases.  As a novel electronic order, the momentum-space instability provides a common thread that links together a variety of previously observed phenomena in BLG, such as spontaneous symmetry breaking and isospin degeneracy lifting.  However, despite its crucial role in defining the interplay between correlation, broken symmetry and isospin order, experimental observation of spontaneous momentum polarization has remained elusive, mostly owing to the lack of viable experimental methods. 

In this work, we show that spontaneous momentum polarization can be directly identified using angle-resolved measurement of second-harmonic nonlinear transport (\arnt). A transport response at the second-harmonic frequency of the AC current bias indicates two-fold rotational symmetry breaking ~\cite{He2022nonlinear,Isobe2020nonlinear,Sinha2022nonlinear,Ma2019nonlinear,Kang2019nonlinear,Sodemann2015nonlinear}. For example, a valley-polarized state, which breaks both two-fold rotational $C_2$ and time-reversal $T$ symmetries, is shown to generate a second-harmonic nonlinear transport response in magic-angle twisted trilayer graphene ~\cite{Zhang2022valley,Zhang2022sunflower}. We propose that the angular dependence of the nonlinear response at the second-harmonic frequency directly reflects the contour of a valley-imbalanced Fermi surface. 

The angle-dependent nonlinear response at the second-harmonic frequency, $V^{2\omega}_{\parallel}(\phi)$, is measured from a BLG sample with the ``sunflower'' geometry (inset in Fig.~\ref{fig1}e, also see Fig.~\ref{figDevice})  ~\cite{Zhang2022valley,Zhang2022sunflower}. $V^{2\omega}_{\parallel}(\phi)$ denotes the voltage response between two contacts aligned parallel to the azimuth direction of current flow $\phi$, which is measured at the second-harmonic frequency with an AC current bias of $I_{AC} = 60$ nA.
Given the key role of two-fold rotational symmetry breaking, we fit the angular dependence of the second-harmonic nonlinear response using a linear combination of one-fold and three-fold symmetric components,
\begin{equation}
    V^{2\omega}_{\parallel}(\phi) = V_1 \textrm{cos}(\phi-\beta_1) + V_3 \textrm{cos}(3(\phi-\beta_3)).
\end{equation}
Here $V_1$ ($V_3$) denotes the amplitude of angular oscillation of the one-fold (three-fold) component. 
A three-fold symmetric Fermi surface of a valley-polarized state, where carriers equally occupy all three Fermi pockets (Fig.~\ref{fig1}c), will give rise to a non-zero $V_3$. The Coulomb-driven instability in the momentum space further breaks the three-fold rotational symmetry $C_3$. The resulting momentum-polarized state is manifested in the one-fold symmetric angular dependence in the nonlinear transport response. As such, $V_1$ offers a direct characterization for the strength of momentum polarization. 
Along the same vein, $\beta_3$ corresponds to the phase of the three-fold oscillation. Whereas $\beta_1$ defines the polar axis of the one-fold component, which is aligned along the direction of the occupied Fermi pocket. Since trigonal-warping-induced Fermi pockets are located in well-defined corners of the momentum space, $\beta_1$ and $\beta_3$ can only take six values that correspond to the azimuth directions of the Fermi pockets. 

These special azimuth angles are determined based on the predominantly three-fold symmetric angular dependence with $V_3 \gg V_1$, as shown in Fig.~\ref{fig1}e. This angular dependence points towards a valley-polarized state, where carriers equally occupy three pockets from valley K' (schematic on the bottom right of Fig.~\ref{fig1}e). The maximum positive nonlinear response defines the azimuth directions of occupied Fermi pockets in valley K', which are labeled as $1'$, $2'$, and $3'$; whereas empty pockets in valley K, labeled as $1$, $2$, and $3$, are located near maximum negative nonlinear response (bottom right panel of Fig.~\ref{fig1}e). In the following, we label the azimuth direction of each Fermi pocket using black dashed lines in polar-coordinate plots. We also mark each azimuth direction with the associated Fermi pocket, in order to identify the carrier occupation of Fermi pockets. In the presence of a momentum-polarized state, we expect a one-fold symmetric nonlinear response with a polar axis aligned along the occupied Fermi pocket.

Unambiguous evidence of spontaneous momentum polarization is revealed by investigating the density-electric-field ($n-D$) map of BLG. Across most of the $n-D$ map, the nonlinear response at the second-harmonic frequency exhibits a predominantly one-fold symmetric angular dependence.  Fig.~\ref{fig2}a plots the polar-axis orientation $\beta_1$ as a function of $n$ and $D$. Most remarkably, the polar axis is always aligned along the direction of a Fermi pocket. As shown in Fig.~\ref{fig2}b-c, the maximum positive (or negative) nonlinear response always occurs near the black dashed lines in the polar-coordinate plots, which is in excellent agreement with the expected behavior of momentum polarization. 
Based on the value of $\beta_1$, the $n-D$ map divides into distinct regimes marked by $A$, $B$, $A'$, $B'$ and $C$.   
The transition boundary between $A$ and $B$ ($A'$ and $B'$) shifts to higher carrier density with increasing $D$. This gives rise to a fan-like diagram that is characteristic of momentum polarization ~\cite{Dong2021momentum}. 

Across the boundary between $A$ and $B$, as well as $A'$ and $B'$, the polar axis exhibits a rotation of $180^{\circ}$ (panel \textit{i} and \textit{v} of Fig.~\ref{fig2}b and Fig.~\ref{fig2}c). The $180^{\circ}$ rotation indicates that charge carriers occupy Fermi pockets from different valleys on opposite sides of the transition. In a small density regime near this transition boundary, $V_1$ diminishes (Fig.~\ref{fig2}e), giving rise to a predominantly three-fold symmetric angular dependence (panel \textit{iii} of Fig.~\ref{fig2}b). This angular dependence points towards a momentum-unpolarized state, MUP, where carriers equally occupy three pockets in the same valley. Interestingly, every transition boundary between regimes with distinct $\beta_1$ is accompanied by a narrow density range of MUP.  Since a polar axis is ill-defined for the MUP phase, its density regimes are marked using the color grey with white solid contours (Fig.~\ref{fig2}a). We label the transition boundary between $A$ and $B$ as \textit{I} for simplicity.
That a MUP phase occurs between momentum polarization in opposite valleys is in excellent agreement with the scheme that Fermi pockets are occupied sequentially with increasing carrier density (Fig.~\ref{fig1}d). According to sequential occupation, regime $A$ corresponds to a fully momentum-polarized state, MP, where all charge carriers occupy a single Fermi pocket. Whereas carriers occupy multiple Fermi pockets across opposite valleys in regime $B$, giving rise to a partially polarized state in both valley and momentum channels, PVP/PMP. For instance, the polar-coordinate plot in panel \textit{v} of Fig.~\ref{fig2}b corresponds to the Fermi pocket occupation in the bottom panel. While all Fermi pockets are occupied, extra carriers condense into pocket $2$ of valley K, which defines the polar axis orientation of the momentum polarization. The strength of momentum-polarization on two sides of transition \textit{I} is further confirmed by the density dependence of $V_1$ across the boundary. As shown in Fig.~\ref{fig2}e, the fully momentum-polarized state on the low-density side of the transition exhibits a $V_1$ that is orders of magnitude larger compared to the partially polarized state on the high-density side. 

A closer look at the transition boundary between $A$ and $B$ reveals a few intermediate steps, as shown in Fig.~\ref{fig2}b. Panels \textit{i} through \textit{v} demonstrate the full evolution of Fermi pocket occupation with increasing charge carrier density across the transition boundary. Adding charge carriers to the momentum-polarized (MP) state in panel \textit{i} gives rise to the partial occupation of pocket $1'$ and $3'$ (panel \textit{ii}). The mixed angular dependence in panel \textit{ii} indicates a fully valley-polarized, but partially momentum-polarized phase in K' valley, VP/PMP (also see Fig.~\ref{1and3fold}c-d).  Similarly, adding charge carriers to the momentum-unpolarized  (MUP) phase in panel \textit{iii} leads to a partial occupation of pocket $2$ and $3$ in valley K, which corresponds to a mixed angular dependence in panel \textit{iv}. Fig.~\ref{fig2}d-e plots the density dependence of $\beta_1$ and $V_1$ across this boundary. Red arrows mark the density corresponding to each polar-coordinate plot in Fig.~\ref{fig2}b. Throughout two transition boundaries, marked by rotations in $\beta_1$ and minima in $V_1$, the polar axis of momentum-polarized state is shown to always align along one of the Fermi pockets. 

%Further increasing carrier density gives rise to the Fermi pocket occupation as shown in panel \textit{v}, which corresponds to the \textit{PVP/PMP} phase in regime $B$. 

It is worth pointing out that a momentum-polarized state breaks in-plane rotational and time-reversal symmetries simultaneously. The prominent nonlinear response observed near the CNP at $D=0$ (Fig.~\ref{fig2}f) indicates that momentum-polarization is likely responsible for previous observations of reduced rotational symmetry, and broken time-reversal symmetry in this regime ~\cite{Weitz2010BLG,Mayorov2011BLG}. Given the unique symmetry requirement, momentum polarization is distinct from other mechanisms for generating the second-harmonic nonlinear transport response, such as Berry-curvature dipole and skew scattering, which relies on sublattice symmetry breaking ~\cite{He2022nonlinear,Isobe2020nonlinear,Sinha2022nonlinear,Ma2019nonlinear,Kang2019nonlinear,Sodemann2015nonlinear}. This distinction is further illustrated by the lack of $D$ dependence in the nonlinear transport response. As shown in Fig.~\ref{fig2}a and Fig.~\ref{figDSI}, the angular dependence of the nonlinear response remains mostly the same upon reversing the electric field $D$. Since the sublattice polarization switches sign upon the reversal of electric field $D$, the lack of dependence on $D$ provides a strong indication that sub-lattice and layer polarization are of secondary importance in the observed nonlinear response. 
Spontaneous momentum polarization is also a distinct order compared to electronic nematicity. An orthorhombic anisotropy preserves both two-fold rotational and time-reversal symmetry and thus does not generate a second-harmonic nonlinear signal. As such, a momentum-polarized state does not directly couple to uniaxial strain in the sample. We note that the second-harmonic nonlinear response along the current flow direction is equivalent to the diode-like nonreciprocity in the DC transport ~\cite{Zhang2022valley}. To distinguish with nematicity, we refer to the underlying electronic order as diodicity, which is defined by simultaneously breaking two-fold rotational and time-reversal symmetries.

\begin{figure*}
\includegraphics[width=0.9\linewidth]{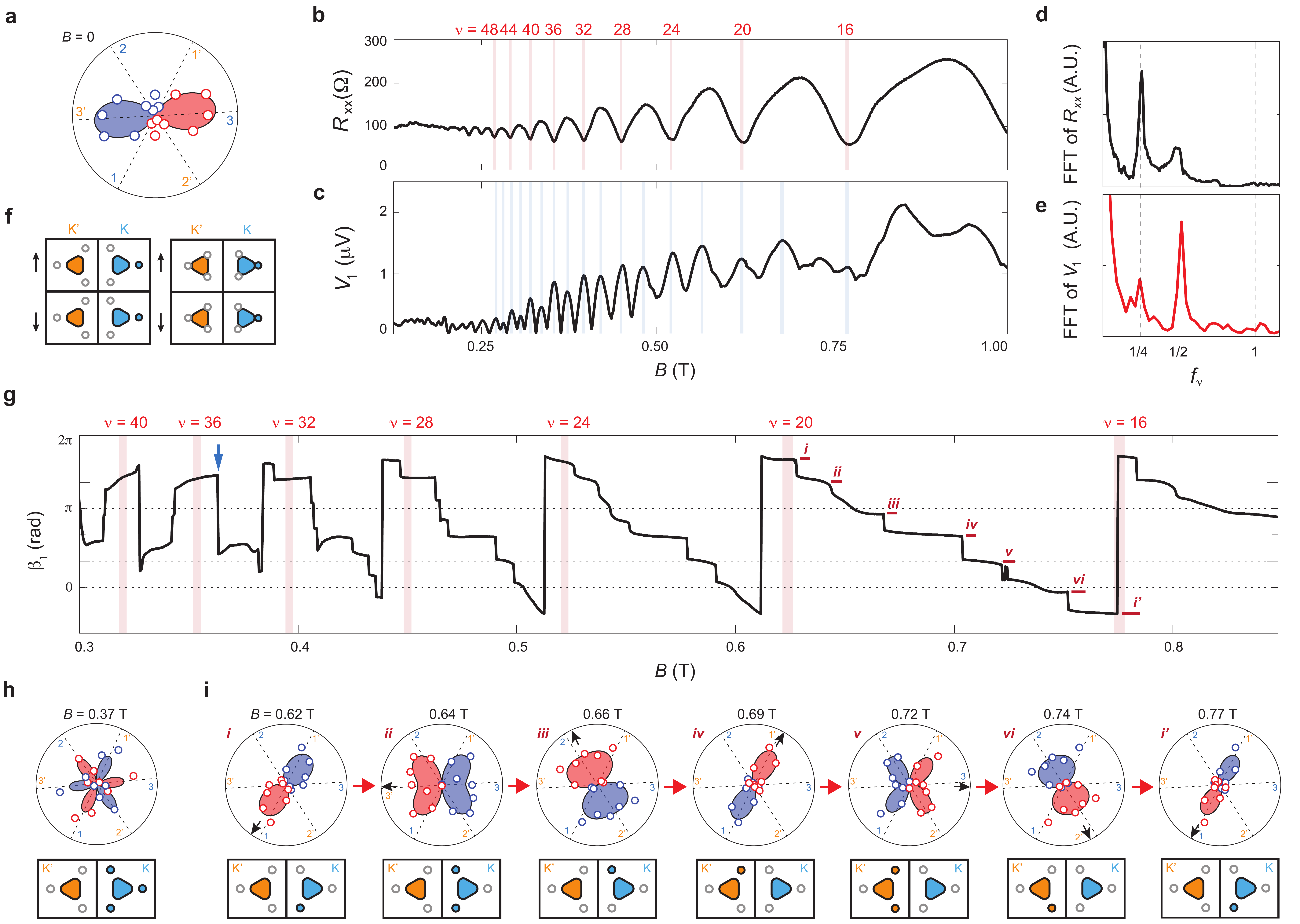}
\caption{\label{figB}{\bf{Tuning momentum-space instability with an out-of-plane $B$-field.} } (a) The polar-coordinate plot of nonlinear transport response at $B=0$.  (b) Longitudinal resistance $R_{\parallel}$ measured at the first-harmonic frequency  and (c) the amplitude of nonlinear response $V_1$ as a function of an out-of-plane magnetic field $B$ measured at $n=-0.3 \times 10^{12}$ cm$^{-2}$ and $D = 0$. Red vertical stripes mark minima in $R_{xx}$, which correspond to the emergent Landau level gaps. Blue vertical stripes mark maxima in $V_1$. (d-e) Fast Fourier transformation of (d) $R_{xx}$ and (e) $V_1$. X-axis is the frequency renormalized to carrier density $n$. (f) Schematic diagram of possible Fermi surface contours. Triangles indicate large, isospin-degenerate Fermi surfaces, whereas circles represent small, trigonally-warped Fermi pockets.  (g) The $B$-dependence of the polar axis orientation $\beta_1$. Red vertical  stripes label the location of minima in $R_{xx}$.  Horizontal red bar marks plateaus in $\beta_1$ between LL fillings $16 < \nu_{LL} <20$. (h-i) The polar-coordinated plots of nonlinear response measured at (h) the transition marked by the blue vertical arrow in panel (g), and (i) each $\beta_1$ plateau  marked with \textit{i} through $vi$ in panel (g). 
 }
\end{figure*}

\begin{figure*}
\includegraphics[width=1\linewidth]{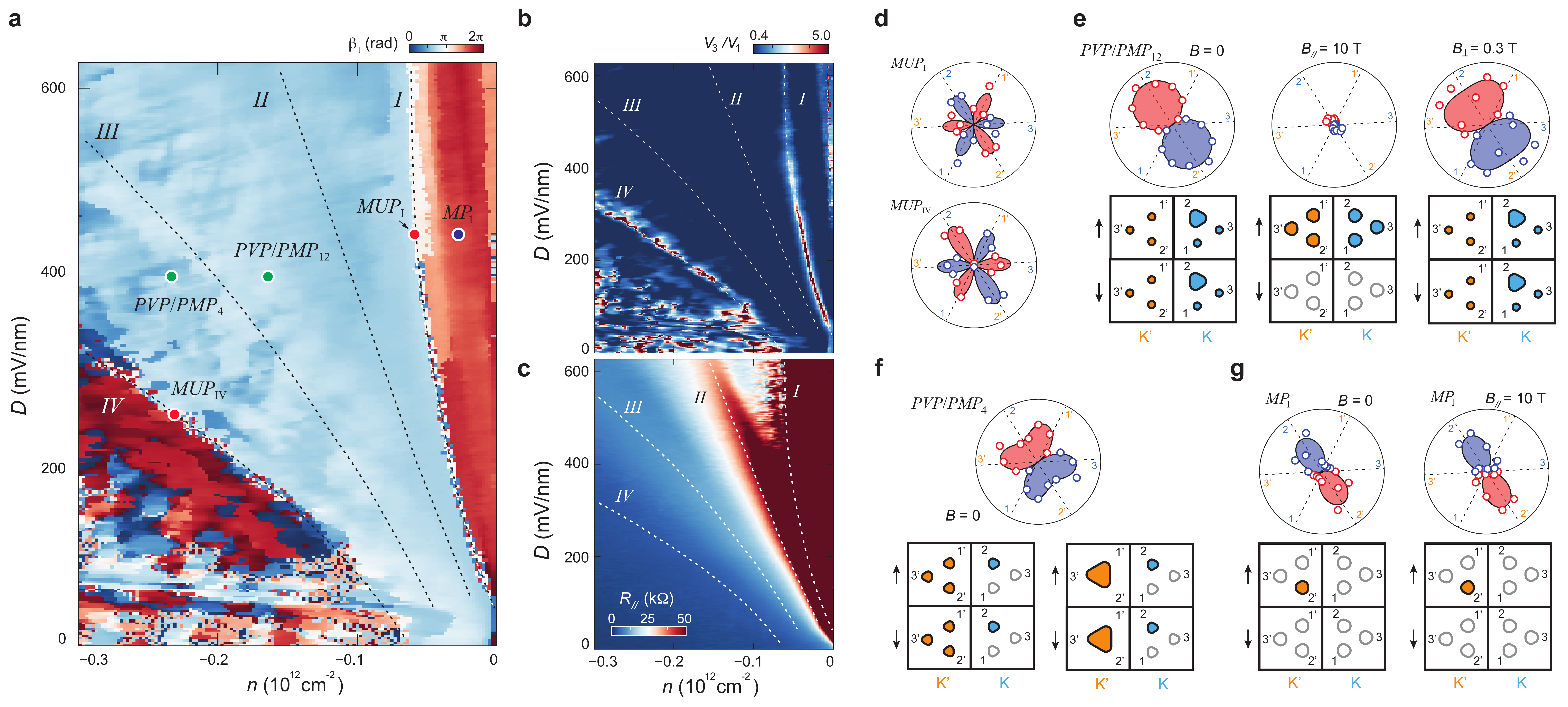}
\caption{\label{fig4}{\bf{Intertwined orders between momentum and isospin polarization.} } (a) Polar-axis orientation $\beta_1$, (b) the ratio of $V_3/V_1$, and (c) longitudinal resistance \Rxx\  as a function of carrier density $n$ and displacement-field $D$. 
(d) Polar-coordinate plots of MUP phases measured along transition boundaries \textit{I} and \textit{IV}. 
(e) Polar-coordinate plots of the PVP/PMP$_{12}$ phase in the regime between \textit{II} and \textit{III}. At $B=0$ (left panel), the angle-dependent nonlinear response is consistent with a partially valley- and momentum-polarized state with $12$ Fermi pockets. Applying a large in-plane magnetic field $B_{\parallel} = 10$ T suppresses the nonlinear response (middle panel) by stabilizing a valley-balanced state. An out-of-plane magnetic field  $B_{\perp} = 0.3$ T has little impact on the nonlinear response (right panel). (f) Polar-coordinate plots of the PVP/PMP$_{4}$ phase in the regime between \textit{III} and \textit{IV}. The bottom panels plot two possible configurations of Fermi surface contour. (g) Polar-coordinate plots of the MP$_{1}$ phase on the low-density side of transition \textit{I}. The nonlinear response in this regime remains mostly the same between $B=0$ (left panel) and  $B_{\parallel} = 10$ T (right panel). 
 }
\end{figure*}

According to the $n-D$ map, the high-density boundary of the fully momentum-polarized state, $MP$, occurs around $n = -0.06 \times 10^{12}$ cm$^{-2}$ at $D = 300$ mV/nm. This is the same order of magnitude compared to the estimated density range where the three-pocket model is applicable ~\cite{Dong2021momentum}. 
In the following, we show that the momentum space instability persists outside of this density regime in the presence of a large Fermi surface (Fig.~\ref{figB}). At $n = -0.3\times 10^{12}$ cm$^{-2}$ and $D=0$, Fig.~\ref{figB}a shows a one-fold symmetric angular dependence in the nonlinear transport response measured at $B=0$, which indicates a momentum-polarized state with the polar axis aligned along Fermi pocket $3$ (Fig.~\ref{figB}a). The underlying Fermi surface contour is revealed by comparing the magneto-oscillation between the longitudinal resistance measured at the first-harmonic frequency, $R_{xx}$, and the parameter $V_1$, which is extracted by fitting the angular dependence of the nonlinear response using Eq.~1. Up to $B=1$ T, the sequence of quantum oscillation in \Rxx\ is predominantly four-fold degeneracy (Fig.~\ref{figB}b), which is manifested in a $f_{\nu}=1/4$ peak in the FFT of $R_{xx}$ (Fig.~\ref{figB}d) ~\cite{Zhang2009,Feldman2009BLG,Maher2014BLG}. The four-fold degeneracy points towards a large, isospin-degenerate Fermi surface. The FFT of $R_{xx}$ also reveals a smaller peak at $f_{\nu}=1/2$, indicative of a small distortion in the Fermi surface. 

The nature of this distortion is revealed by examining the magneto-oscillation of $V_1$. Within each oscillation of $R_{xx}$ (marked by red vertical stripes in Fig.~\ref{figB}b), $V_1$ exhibits two maxima (blue vertical stripes in Fig.~\ref{figB}c) and two minima. Such a doubling in the frequency is reflected by a prominent peak at $f_{\nu}= 1/2$ in the FFT of $V_1$. 
Combined, our findings suggest that the $f_{\nu}=1/4$ peak results from a large Fermi surface that is mostly four-fold degenerate, whereas the $f_{\nu}=1/2$ peak arises from a small distortion in the Fermi surface that is spin-degenerate but valley-imbalanced. Fig.~\ref{figB}f plots two possible Fermi surface contours, with triangles denoting the large, four-fold degenerate Fermi surfaces and circles representing small Fermi pockets.  The scenario where each isospin quadrant consists of four Fermi surfaces has been proposed by prior discussions ~\cite{McCann2006,McCann2013BLGreview}. In this scenario, the momentum-polarized state at $B=0$ naturally results from the exchange-driven instability among the small Fermi pockets. Alternatively, a net momentum polarization could arise from an exchange-induced distortion, which induces extra charge carriers concentrated near a corner of the large Fermi surface ~\cite{Jung2015momentum}. This is captured by the schematic diagram in the right panel of Fig.~\ref{figB}f, where small Fermi pockets are connected to the large surface. While the \arnt\ provides identification for the polar axis of momentum polarization, these possible scenarios in Fig.~\ref{figB}f cannot be distinguished based on \arnt\ alone. In the following, we use the schematic with separated pockets to denote the polar axis orientation of momentum polarization, without making a claim regarding the radial location of small Fermi pockets.

The magneto-oscillation in $V_1$ shed light on an intriguing interplay between the Coulomb-driven instability in the momentum space and the out-of-plane $B$. As shown in Fig.~\ref{figB}g, this interplay induces a cascade of transitions in the polar axis orientation $\beta_1$ with varying $B$. 
In the range of $0.25< B < 0.4$ T, the polar axis rotates twice within each isospin-degenerate Landau level $LL$. Each rotation is accompanied by a three-fold symmetric angular dependence in the nonlinear response, as shown in the polar-coordinate plot of Fig.~\ref{figB}h (also see Fig.~\ref{figB2fold}). This behavior is characteristic of a simultaneous transition in the valley- and momentum-polarization. According to Fig.~\ref{figB}g, two transitions in the valley isospin order occurs within each LL.  Since \Rxx\ shows no sign of quantum Hall ferromagnetism~\cite{Young2012ferromagnetism}, we conclude that these valley polarization transitions are gapless. This points towards an incipient isospin ferromagnetic order arising from the momentum-space instability among the small Fermi pockets. As such,  the momentum space instability provides a direct link between isospin degeneracy lifting, the reduced rotational symmetry, and the broken time-reversal symmetry. 

The application of the $B$-field provides a uniquely efficient knob for tuning the momentum space instability (Fig.~\ref{figB}i), evidenced by a series of well-defined plateaus in the polar axis orientation upone further increasing $B$. 
The horizontal red line in Fig.~\ref{figB}g marks the position of six plateaus in the filling range $16 < \nu_{LL} < 20$. According to the angular dependence of the nonlinear response, each plateau in $\beta_1$ corresponds to a momentum-polarized state. Adjacent plateaus correspond to a $60^{\circ}$ rotation in the polar axis of momentum polarization. As such, the polar axis rotates by a full $360^{\circ}$ across an isospin-degenerate Landau level, realizing all possible configurations of the momentum-polarized order. 
That all possible polar axis orientations are accessed by varying $B$ is a strong indication of spontaneous rotational symmetry breaking, which is driven by the momentum space instability. As such, the potential influence of stacking boundaries and topological defects in BLG is secondary at best ~\cite{Martin2008BLGsoliton,Alden2013BLGsoliton,Ju2015BLGsoliton}.
$B$-induced tunability is also observed in the low-density regime where charge carriers occupy separate Fermi pockets (Fig.~\ref{figBSI}). Since the Fermi surface contour is highly sensitive to the $B$-field, it puts some constraints on our ability to characterize the Fermi surface contour based on magneto-oscillation alone.

Having established the method to probe and characterize momentum polarization, we are now in a position to investigate its connection with previously observed isospin-degeneracy lifting at a large electric field ~\cite{Zhou2022BLG,Zhang2022BLG,De2022BLG}. 
Fig.~\ref{fig4}a-c plots transport properties measured over a quadrant of the $n-D$ map with hole-type charge carriers and $D > 0$. Since a rotation in the polar axis is usually accompanied by a suppression in $V_1$, the location of the transition coincides with a prominent peak in the ratio of $V_3/V_1$. Fig.~\ref{fig4}b plots $V_3/V_1$ across the $n-D$ map, which highlights the location of the transition boundaries as red color regimes in the chosen color scale.
Based on the angular dependence of the nonlinear transport response (see Fig.~\ref{figAllPolars}), we identify four transition boundaries in the $n-D$ map, which are labeled \textit{I} to \textit{IV}. \textit{I} is the same boundary between regime $A$ and $B$ as shown in Fig.~\ref{fig2}a, extended to a larger range in $n$ and $D$. Similar to \textit{I}, \textit{IV} is accompanied by a large rotation in the polar axis (Fig.~\ref{fig4}a), which coincides with a small density regime with three-fold symmetric angular dependence in the nonlinear response.  The three-fold symmetric response is manifested in an enhanced ratio $V_3/V_1$, which is marked by red color in the chosen color scale in Fig.~\ref{fig4}b. Both \textit{I} and \textit{IV}   correspond to simultaneous transitions in the valley and momentum polarization. Notably, the three-fold symmetric angular dependence near boundary \textit{I} and \textit{IV} exhibit opposite polarities (Fig.~\ref{fig4}d). This points towards momentum-unpolarized phases with opposite valley isospin orders, which are labeled as MUP$_I$ and MUP$_{IV}$, respectively. 

Unlike \textit{I} and \textit{IV}, boundaries \textit{II} and \textit{III} are not associated with rotations in the polar-axis or three-fold symmetric angular dependence. Throughout the density regime between \textit{I} and \textit{IV}, the angle-dependent nonlinear response exhibits a prominent one-fold component, with the polar axis aligned along Fermi pocket $2$. As shown in Fig.~\ref{figAllPolars} and Fig.~\ref{figRegimeB}, \textit{II} and \textit{III} separate regimes of the $n-D$ map with slightly different angular symmetry. Between \textit{II} and \textit{III}, the angular dependence is predominantly one-fold symmetric; whereas the regimes between \textit{I} and \textit{II}, as well as \textit{III} and \textit{IV}, are best fit with a linear combination of one-fold and three-fold components in the angle-dependent nonlinear response. 

The angular dependence of the nonlinear response allows us to deduce the Fermi pocket occupation. On the high-density side of \textit{III}, the one-fold and three-fold components of the angular dependence exhibit opposite polarities. This implies that the Fermi surface contour is distinct across opposite valleys (also see Fig.~\ref{1and3fold}). The one-fold symmetric component arises from the momentum-space instability in valley K, which induces charge carriers to condense into pocket $2$; whereas the Fermi surface contour in valley K' is invariant under a three-fold rotation $C_3$, generating the three-fold symmetric component in the angular dependence. The $C_3$-preserving Fermi surface in valley K' could result from two possibilities: a large Fermi surface or three separate but equally-occupied pockets (the bottom panels of Fig.~\ref{figB}f). On the other hand, the low-density side of \textit{III} is best explained by the scenario where charge carriers occupying $12$ Fermi pockets. While pockets $1$, $1'$, $3$, and $3'$ are equally occupied, the momentum-space instability gives rise to a prominent imbalance between pocket $2$ and $2'$, which is responsible for the momentum polarization and the associated one-fold symmetric angular dependence in the nonlinear response. 

The trajectories of \textit{II} and \textit{III} in the $n-D$ map show excellent agreement with the peak position in the longitudinal resistance measured with the first-harmonic frequency (Fig.~\ref{fig4}c) ~\cite{Zhou2022BLG,Zhang2022BLG,De2022BLG}. Notably, the angular dependence of the nonlinear transport response is mostly consistent with the Fermi surface contours extracted from the magneto-oscillation from previous observations  ~\cite{Zhou2022BLG}. On the low-density side of \textit{III}, magneto-oscillation  points towards $12$ equally occupied Fermi pockets across four isospin quadrants. Since the momentum space instability couples strongly with an out-of-plane $B$-field, as shown in Fig.~\ref{figB}g, we conjecture that the magneto-oscillation is insensitive to the momentum-space instability at $B=0$, which creates an imbalance between pocket $2$ and $2'$. On the high-density side of \textit{III}, both magneto-oscillation and \arnt\ are consistent with two large and two small Fermi surfaces, as shown in the bottom right panel of Fig.~\ref{fig4}f. Since the entire regime between \textit{I} and \textit{IV} is partially valley- and momentum-polarized, we will label the ground state based on the number of Fermi pockets. For instance, the regime between \textit{II} and \textit{III} (\textit{III} and \textit{IV}) is occupied by the PVP/PMP$_{12}$ (PVP/PMP$_4$) phase. 
 
While the angular dependence of the nonlinear response is not directly sensitive to the spin order, the application of an in-plane magnetic field \Bpara\ offers insights into the spin degrees of freedom. In the presence of a large in-plane magnetic field of \Bpara $=10$ T, the fully momentum-polarized (MP) phase on the low density side of boundary \textit{I} remains mostly the same as $B=0$ (Fig.~\ref{fig4}g). The insensitivity to \Bpara, combined with the angle-dependent nonlinear response, suggests that charge carriers occupy one Fermi pocket that is fully polarized across spin, valley isospin and momentum channels. We will refer to this phase as MP$_{1}$. In stark contrast with MP$_{1}$, both PVP/PMP phases are highly dependent on \Bpara.  As shown in the middle panel of Fig.~\ref{fig4}e, the application of \Bpara $=10$ T fully suppresses the nonlinear response in the PVP/PMP$_{12}$ phase  (see also Fig.~\ref{figBinplane}). In comparison, an out-of-plane $B$-field has little impact (right panel of Fig.~\ref{fig4}e). This confirms that PVP/PMP phases are spin degenerate at $B=0$. As a large \Bpara\ lifts the spin degeneracy, the momentum space instability is suppressed, resulting in a valley-balanced Fermi surface occupation with a diminishing nonlinear response at the second-harmonic frequency.

That a Zeeman-induced spin order suppresses valley and momentum polarization points towards a competition in the exchange-driven instability between the spin and orbital channels ~\cite{Kang2019IsospinFerromagnet}. This competition is particularly intriguing since the PVP/PMP phases are directly linked to the superconducting phase, which is stabilized in the presence of an in-plane magnetic field or proximity with a tungsten diselenide crystal ~\cite{Zhou2022BLG,Zhang2022BLG}. Our observation raises the possibility of an interesting interplay between the stability of the superconducting phase and the momentum-space instability ~\cite{Wagner2023momentum,Dong2022triplet,Curtis2022triplet,Jimeno2022momentum}. 
Furthermore, the universal presence of momentum polarization across the phase space of BLG suggests that the Coulomb-driven instability in the momentum space is fundamental to the electronic order in multilayer graphene systems. A better understanding of momentum degrees of freedom could hold the key to unraveling the nature of other emergent phenomena in this system.
Beyond the identification of spontaneous momentum polarization, our findings also establish the \arnt\ as a highly sensitive tool for resolving spontaneously broken symmetries in multi-layer graphene systems.

\section*{Acknowledgments}
We wish to express sincere gratitude to Daniel Mark, Leonid Levitov, Oskar Vafek, Andrea Young, Zhiyu Dong, and Dmitry Chichinadze for helpful discussions. J-X.L. and J.I.A.L. acknowledge funding from NSF DMR-2143384. Device fabrication was performed in the Institute for Molecular and Nanoscale Innovation at Brown University.
K.W. and T.T. acknowledge support from the Elemental Strategy Initiative
conducted by the MEXT, Japan (Grant Number JPMXP0112101001) and  JSPS
KAKENHI (Grant Numbers 19H05790, 20H00354 and 21H05233). The work at Massachusetts Institute of Technology was supported by a Simons Investigator Award from the Simons Foundation.

\bibliography{Li_ref}

\newpage

\newpage
\clearpage

\pagebreak
\begin{widetext}
\section{Supplementary Materials}

\begin{center}
\textbf{\large Spontaneous momentum polarization and diodicity in Bernal bilayer graphene}\\
\vspace{10pt}
Jiang-Xiazi Lin, Yibang Wang,
Naiyuan J. Zhang,
 Kenji Watanabe, Takashi Taniguchi, Liang Fu and J.I.A. Li$^{\dag}$

\vspace{10pt}
$^{\dag}$ Corresponding author. Email: jia$\_$li@brown.edu
\end{center}

%\noindent\textbf{This PDF file includes:}

%\noindent{Materials and Methods}

%\noindent{Supplementary Text}

%\noindent{Figs. S1 to S9}

\renewcommand{\vec}[1]{\boldsymbol{#1}}

\renewcommand{\thefigure}{S\arabic{figure}}
\def\theequation{S\arabic{equation}}
\def\thetable{S\Roman{table}}
\setcounter{figure}{0}
\setcounter{equation}{0}

%\newpage

%\newpage

\section{Supplementary Text}

\subsection{The Angle-Dependence of the Second-Harmonic Nonlinear Transport Response}

\begin{figure*}[!b]
\includegraphics[width=1\linewidth]{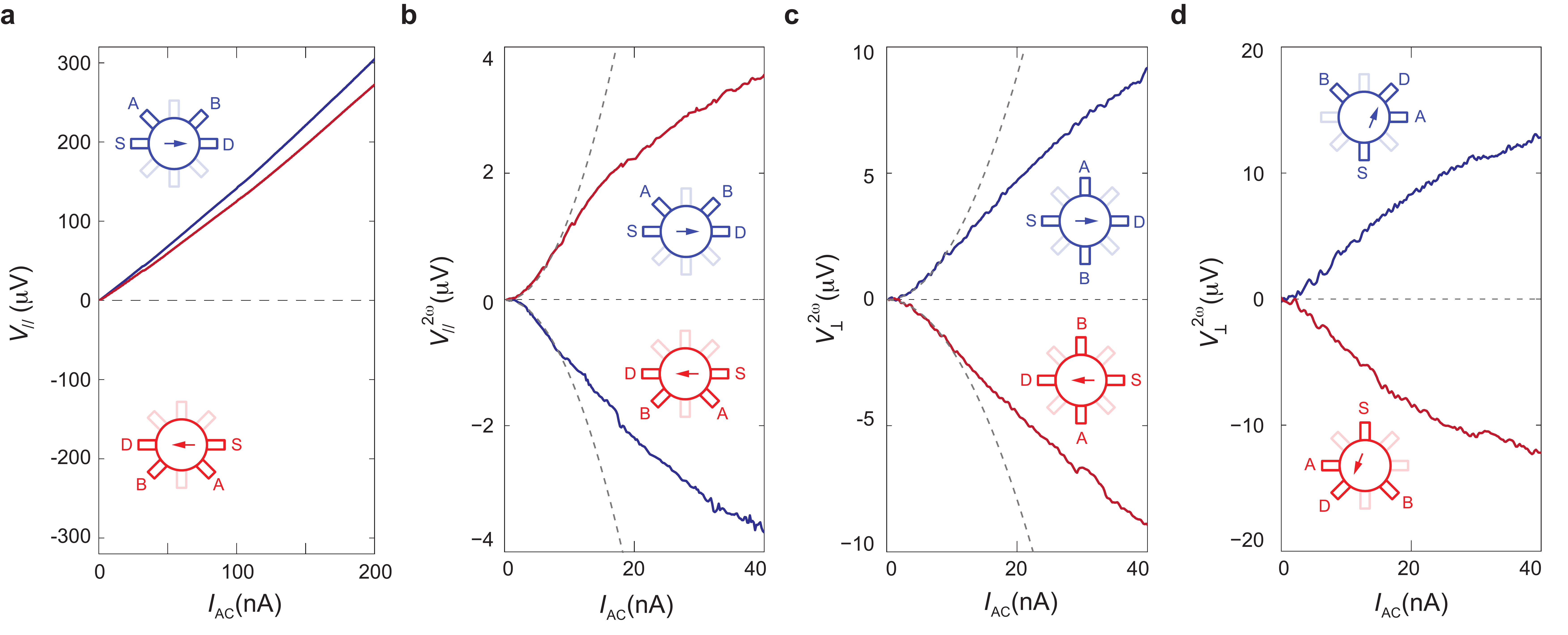}
\caption{\label{fig180}{\bf{The Second-Harmonic Nonlinear Transport Response.} } (a) Current-Voltage (I-V) characteristic of linear transport response, which is defined as the voltage response measured at the first-harmonic frequency of the AC current bias. $V_{\parallel}$ denotes the voltage response measured between two contacts that are parallel to the current flow direction. Similarly, we will use $V_{\perp}$ to mark the voltage response between two contacts perpendicular  to the current flow direction. Blue and red insets show two different measurement configurations, which are related by a $180^{\circ}$ rotation. Measured at the first-harmonic frequency, the linear transport response is mostly invariant under the two-fold inplane rotation of the measurement setup.  (b-d) I-V curves of the second-harmonic nonlinear transport response measured with different current flow directions. $V_{\parallel}^{2\omega}$ and $V_{\perp}^{2\omega}$ denote the second-harmonic nonlinear response between two contacts that are parallel and perpendicular  to the current flow direction, respectively. (b) Blue (red) curve corresponds to the current-dependence of $V_{\parallel}^{2\omega}$ with current flowing along $\phi = 0^{\circ}$ ($\phi = 180^{\circ}$). (c) Blue (red) curve corresponds to the current-dependence of $V_{\perp}^{2\omega}$ with current flowing along $\phi = 0^{\circ}$ ($\phi = 180^{\circ}$). (d) Blue (red) curve corresponds to the current-dependence of $V_{\perp}^{2\omega}$ with current flowing along $\phi = 67.5^{\circ}$ ($\phi = 247.5^{\circ}$). Dashed grey lines in panels (c-d) denote a quadratic dependence on current. The IV curves show good agreement with the quadratic fit at a small current while deviating from the quadratic fit at $I_{AC} > 20$ nA. All measurements are performed at $n=-0.16 \times 10^{12}$ cm$^{-2}$, $D = 652$ mV/nm, $T=20$ mK and $B = 0$.
 } 
\end{figure*}

\begin{figure*}[h]
\includegraphics[width=0.77\linewidth]{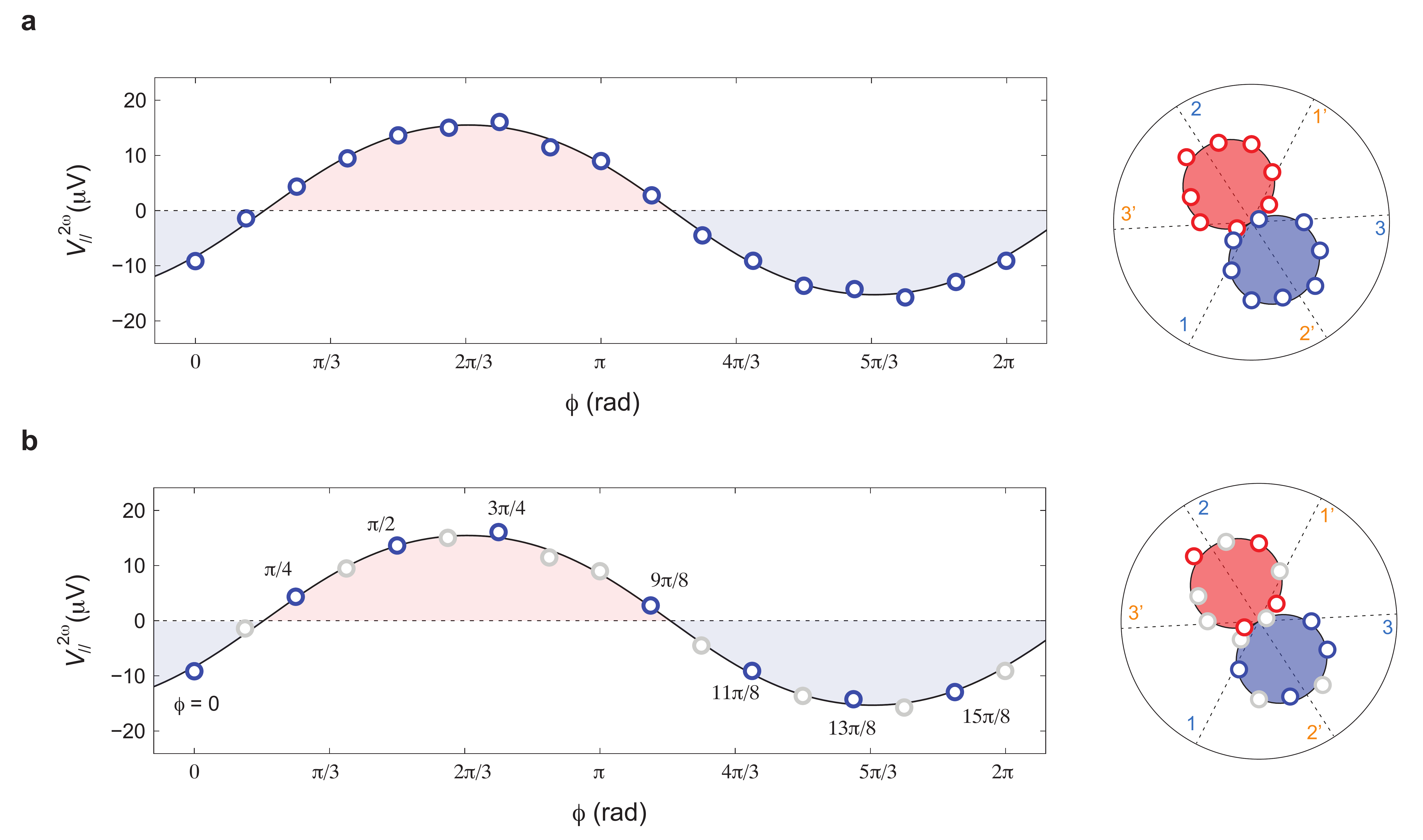}
\caption{\label{fig16}{\bf{The $8$-point method of \arnt.}} (a) The angle-dependence of $V_{\parallel}^{2\omega}$  measured at all $16$ azimuth directions of current flow. The black solid line is the best fit to all data points using a one-fold symmetric function. Right panel shows the polar-coordinate plot of the same data. (b) Blue circles mark the same data points in panel (a) along eight azimuth angles, $\phi = 0$, $\pi/4$, $\pi/2$, $3\pi/4$, $9\pi/8$, $11\pi/8$, $13\pi/4$, and $15\pi/8$. Grey circles are extracted based on blue data points using Eq.~S1. Panels (a) and (b) ($8$-point method) give rise to the same angular fit. The measurement is performed in the PVP/PMP$_{12}$ phase. 
 }
\end{figure*}

\begin{figure*}[!b]
\includegraphics[width=0.77\linewidth]{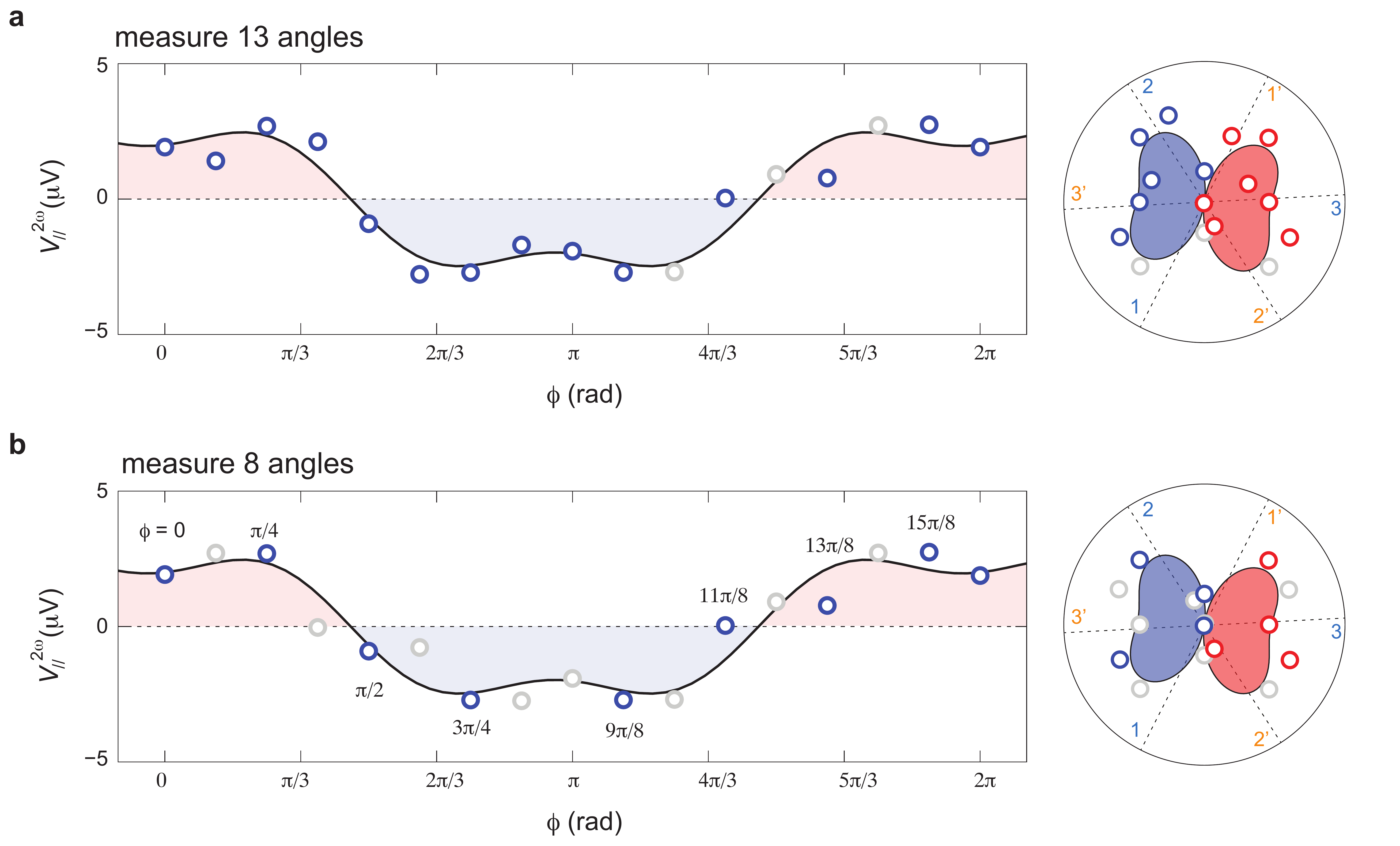}
\caption{\label{fig13} (a-b) shows the angle-resolved measurement of nonlinear response measured at $n = -0.31 \times 10^{12}$ cm$^{-2}$ and $D = 284$ mV/nm, which is on the low-density side of transition boundary \textit{I}. The measurement in panel (a) is performed with the current flowing along $13$ azimuth angles (blue circles). The measurement in panel (b) is performed with current flowing along $8$ azimuth angles. Grey circles are nonlinear response values extracted based on Eq.~S1. Two measurements return the same angular fit for the nonlinear response. The consistency between these two methods confirms the validity of the $8$-point measurement method. 
 }
\end{figure*}

Fig.~\ref{fig180} demonstrates the I-V characteristic of the second-harmonic nonlinear transport response. The ``sunflower'' sample geometry preserves the two-fold in-plane rotational $C_2$ symmetry. Given the electronic state is invariant under a two-fold rotation, the transport response is expected to remain the same when the measurement configuration is ``rotated'' by $180^{\circ}$. That is to say, switch the contact used as source and drain for the current bias, while using the contact on the opposite side of the sample for voltage measurement (see insets of Fig.~\ref{fig180}). Under a $180^{\circ}$ ``rotation'' in the measurement configuration, the linear transport responses measured at the first-harmonic frequency remain mostly the same (Fig.~\ref{fig180}a). The slight difference between blue and red traces likely results from the influence of higher harmonic terms at large current bias. 

An electronic state with spontaneous momentum polarization simultaneously breaks two-fold rotational $C_2$ symmetry and time-reversal $T$ symmetry. Such an odd-parity order is expected to generate a nonlinear transport response at the second-harmonic frequency. The most characteristic signature of this second-harmonic nonlinear response is the sign reversal upon ``rotating'' the measurement configuration by $180^{\circ}$. This sign-reversal has been observed in two types of material systems. In the first type, the generation of the second-harmonic nonlinear response is directly linked to the underlying lattice structure, which breaks the two-fold rotational symmetry. For example, the WTe$_2$ crystal breaks the inversion symmetry along the mirror axis  ~\cite{Kang2019nonlinear,Ma2019nonlinear}. Alternatively, the alignment between graphene and the hBN substrate lifts the sublattice degeneracy of graphene, which results in the two-fold rotational symmetry breaking ~\cite{He2022nonlinear}. In the second type, the second-harmonic nonlinear response arises from the Coulomb-driven polarization in the valley isospin order. In this scenario, the two-fold rotational symmetry is spontaneously broken. A purely Coulomb-driven origin in the second-harmonic nonlinear response is recently reported in magic-angle twisted trilayer graphene ~\cite{Zhang2022valley}.  Regardless of its mechanism, the second-harmonic nonlinear response is expected to exhibit an angular dependence that is either one-fold or three-fold symmetric. The odd parity angular dependence provides the basis for the characteristic sign reversal behavior, which is captured by the following equation,
\begin{equation}
    V_{\parallel}^{2\omega}(\phi) = -V_{\parallel}^{2\omega}(\phi + \pi).
\end{equation}

Fig.~\ref{fig180} shows the sign reversal behavior measured in Bernal-stacked BLG. Upon ``rotating'' the measurement configuration, sign reversal occurs in the nonlinear transport response measured from both longitudinal and transverse channels, \emph{i.e.} voltage response is measured between two contacts aligned parallel (Fig.~\ref{fig180}b) and perpendicular (Fig.~\ref{fig180}c-d) to the direction of current flow. The sign reversal occurs regardless of the azimuth direction of current flow. In most of this work, the nonlinear response at the second-harmonic frequency is measured parallel to the current flow direction. In this scenario, the nonlinear response is directly linked to the diode-like nonreciprocity in the DC IV curve ~\cite{Zhang2022valley}.

The ``sunflower'' sample geometry with $8$ electrical contacts allow us to flow current in $16$ azimuth directions from $0$ to $360^{\circ}$ ~\cite{Zhang2022sunflower}.   Fig.~\ref{fig16}a plots the angle-dependent nonlinear transport response of the PVP/PMP$_{12}$ phase, measured with current flowing along $16$ azimuth directions $\phi$. As a function of azimuth angle $\phi$, $V_{\parallel}^{2\omega}$ exhibits a well-defined one-fold oscillation in the range of $0 < \phi < 360^{\circ}$.  A one-fold symmetric angular dependence is consistent with Eq.~S1. 

Owing to the sign reversal behavior in Eq.~S1, there is redundancy in measuring all $16$ azimuth angles. Fig.~\ref{fig16}b shows a more efficient angle-resolved measurement, where $V_{\parallel}^{2\omega}$ is measured along eight azimuth directions, $\phi = 0$, $\pi/4$, $\pi/2$, $3\pi/4$, $9\pi/8$, $11\pi/8$, $13\pi/4$, and $15\pi/8$. For simplicity, we will refer to this measurement scheme as the $8$-point measurement, which is in contrast with the $16$-point measurement. Based on the second-harmonic nonlinear transport response along these $8$ azimuth directions (blue circles in Fig.~\ref{fig16}b), we can extract the nonlinear response along the other $8$ angles using Eq.~S1, which are shown as grey circles in Fig.~\ref{fig16}b. By comparing Fig.~\ref{fig16}a and b, we show that the $8$-points measurement produces the same angular dependence compared to the $16$-points measurement. Since the $8$-point method covers $4$ angles in the range of $0 < \phi < \pi$ and another $4$ in the range of $\pi < \phi < 2\pi$, fitting these $8$ points alone produces the same angular dependence. As such, the angular dependence extracted from the $8$-point method does not rely on the validity of Eq.~S1. 

While the nonlinear transport responses measured along eight azimuth directions are sufficient to determine the polar axis orientation, a typical angle-dependent measurement, as shown in the main text, often includes more than eight azimuth angles. The addition points offer extra confirmation for the validity of Eq.~S1 (Fig.~\ref{fig13}). Throughout the main text, second-harmonic nonlinear transport measurements are performed with an AC current bias of $60$ nA with a frequency of $13$ Hz.

\subsection{The $n-D$ Map}

\begin{figure*}
\includegraphics[width=1\linewidth]{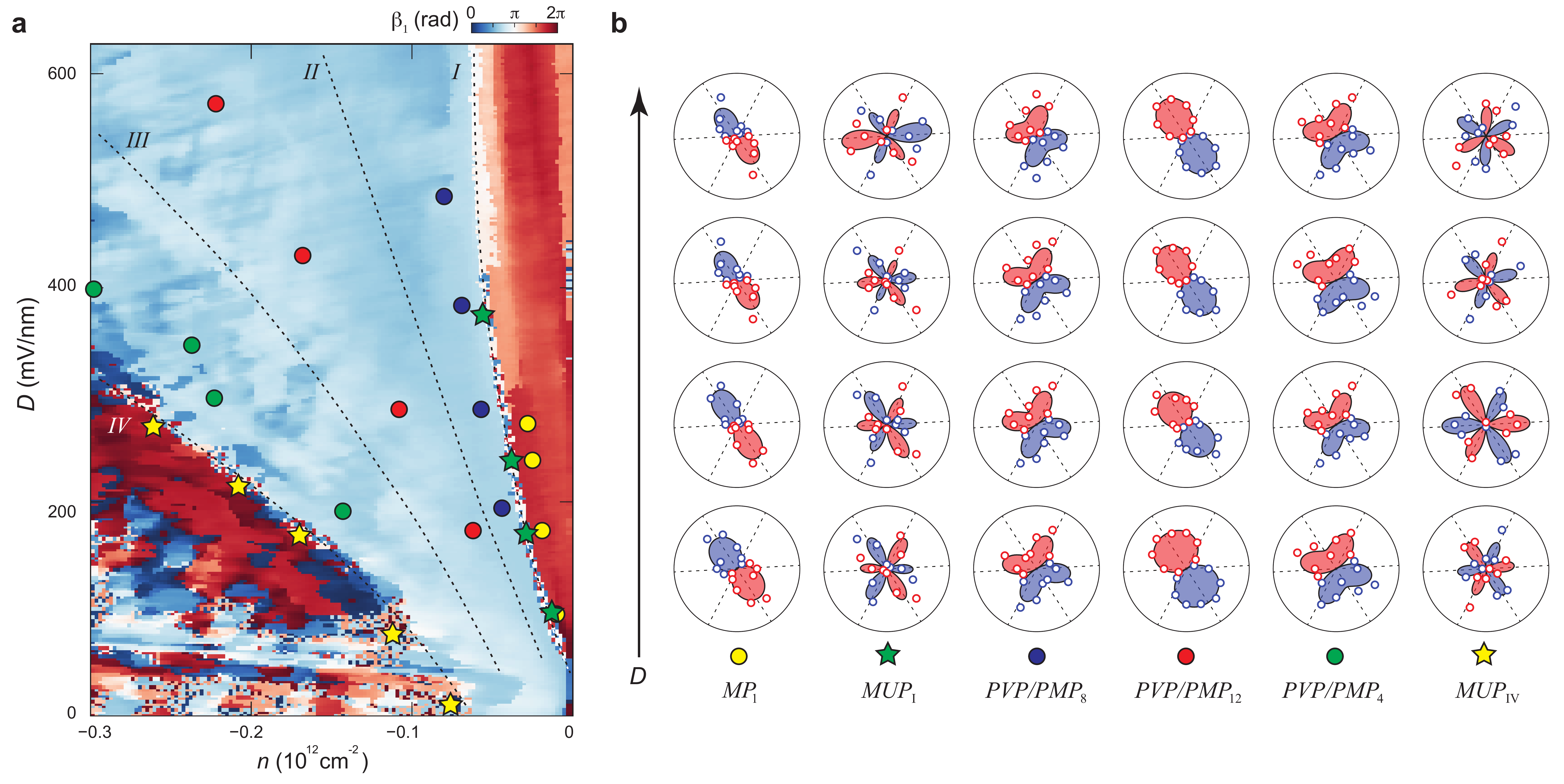}
\caption{\label{figAllPolars}{\bf{Examples of \arnt\ in each isospin-ordered phase.} } (a) Polar-axis orientation $\beta_1$ as a function of carrier density $n$ and displacement-field $D$. (b) Polar-coordinate plots of the \arnt\ measured at the second-harmonic frequency. The colored symbol denotes locations in the $n-D$ map where the angular dependence is measured from. For example, yellow circles are all located on the low density side of \textit{I}. Green and yellow stars mark the transition boundaries \textit{I} and \textit{IV}. Red circles are between boundaries \textit{II} and \textit{III}. $D$ increases in each column from bottom to top. Across most of the $n-D$ map, the angular dependence of both one-fold and three-fold symmetric response are in excellent agreement with the three-pocket model, \emph{i.e.} the polar axis of one-fold symmetric response is aligned along one of six special azimuth directions, while the three-fold symmetric behavior shares the same $\beta_3$. With increasing $n$ and $D$, the angular dependence along the MUP$_{IV}$ phase starts to deviate from the azimuth direction of Fermi pockets (dashed lines in polar coordinate plots). This could arise from the formation of domains of different momentum polarization at large densities. 
 }
\end{figure*}

Fig.~\ref{figAllPolars}a plots the evolution of $\beta_1$ across the same $n-D$ map as shown in Fig.~\ref{fig4}a. Boundaries \textit{I} through \textit{IV} are marked with black dashed lines. The $n-D$ map is divided into $5$ regimes by these boundaries. From low to high density, four different ground states on the low-density side of \textit{IV} are MP$_1$, PVP/PMP$_8$, PVP/PMP$_{12}$, and PVP/PMP$_4$. Polar-coordinate plots in Fig.~\ref{figAllPolars}b show the angular dependence of the nonlinear response measured at different densities $n$ and electric field $D$. According to these polar plots, MP$_1$, PVP/PMP$_8$, PVP/PMP$_{12}$, and PVP/PMP$_4$ remain mostly the same with varying electric $D$. The angular dependence of MUP$_{IV}$ deviates from the expected angles with increasing $D$, which likely results from the formation of domains with different momentum-polarization.

Fig.~\ref{figRegimeB} examines the density dependence of nonlinear response across different regimes. Fig.~\ref{figRegimeB}a shows a characteristic polar plot from each density regime.  Fig.~\ref{figRegimeB}b-c plots the density dependence of $V_1$ and $\beta_1$ measured at a constant electric field value of $D = 140$ mV/nm. According to Fig.~\ref{figRegimeB}b-c, transitions \textit{I} and \textit{IV} are distinct from \textit{II} and \textit{III}. There is no detectable rotation in the polar axis around \textit{II} and \textit{III}. As shown in the polar plots in Fig.~\ref{figRegimeB}a, the one-fold component of the angular dependence points in the same direction for PVP/PMP$_8$, PVP/PMP$_{12}$, and PVP/PMP$_4$ phases. At the same time, $V_1$ does not show prominent dips, indicating that a prominent one-fold component is always present in the density regime between \textit{I} and \textit{IV}. 

The bottom panels in Fig.~\ref{figRegimeB}a plots the schematic diagram for possible configurations of Fermi pocket occupation. The identification of  PVP/PMP$_{12}$ and PVP/PMP$_4$ are discussed in Fig.~\ref{fig4}, which is based on the combination of angle-dependent nonlinear response and the magneto-oscillation measurement. Since PVP/PMP$_{8}$ is on the low-density side of PVP/PMP$_{12}$, we infer that the three-pocket model is applicable for describing the Fermi surface contour in valley K'. As such, charge carriers occupy $8$ Fermi pockets between transition boundaries \textit{I} and \textit{II}.

Among PVP/PMP$_8$, PVP/PMP$_{12}$, and PVP/PMP$_4$ phases, the momentum-space instability occurs in valley K, inducing charge carriers to condense into pocket $2$. Variation in the angular dependence of the nonlinear response arises from carrier distribution amongst other Fermi pockets. For example, adding charge carriers to the PVP/PMP$_8$ phase gives rise to partial occupation in pocket $1$ and $3$. Since carrier density in pocket $2$ remains the highest, the polar axis of momentum polarization remains the same between PVP/PMP$_8$ and PVP/PMP$_{12}$.

The three-fold symmetric response at boundary \textit{I} (MUP$_I$) is robust against both \Bperp (Fig.~\ref{figBSI}d) and \Bpara\ (Fig.~\ref{figBinplane}). As such, we deduce that it is valley-polarized, spin-polarized and momentum unpolarized. On the other hand, the three-fold symmetric state at boundary \textit{IV} is not robust against a magnetic field (Fig.~\ref{figBSI}b). Therefore, we propose that MUP$_{IV}$ is spin unpolarized, partially valley polarized and momentum unpolarized.

\begin{figure*}[!t]
\includegraphics[width=0.85\linewidth]{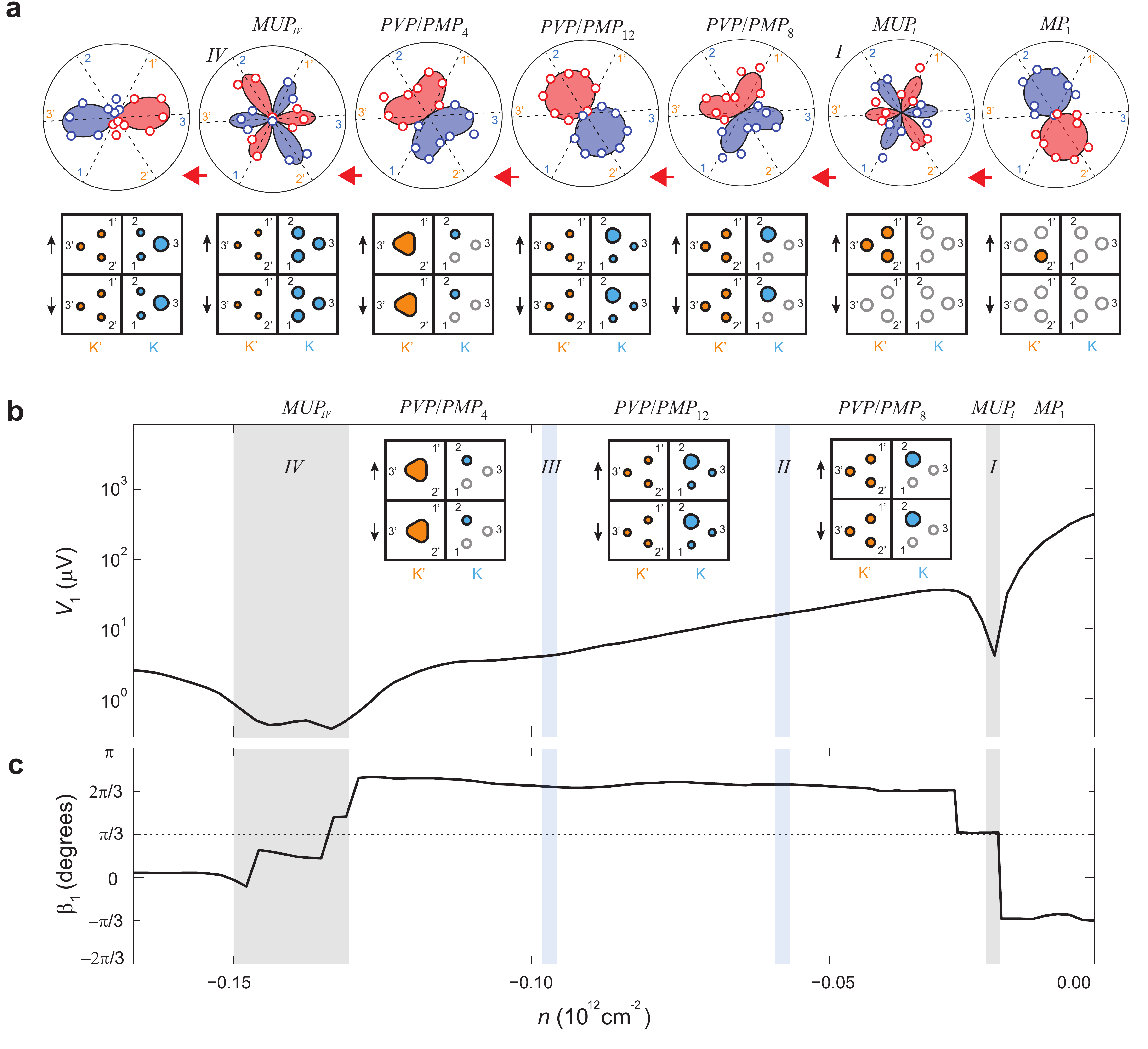}
\caption{\label{figRegimeB}{\bf{Density dependence of \arnt\ across different regimes.} } (a) Polar-coordinate plots of angle-dependent nonlinear response across transition boundaries \textit{I} through \textit{IV}. (b) $V_1$ and (c) $\beta_1$ as a function of charge carrier density measured at $D = 140$ mV/nm. Transition \textit{I} and \textit{IV} coincide with a rotation in the polar axis, which is accompanied by a $C_3$-preserving angle-dependent nonlinear response. 
 }
\end{figure*}

\begin{figure*}
\includegraphics[width=1\linewidth]{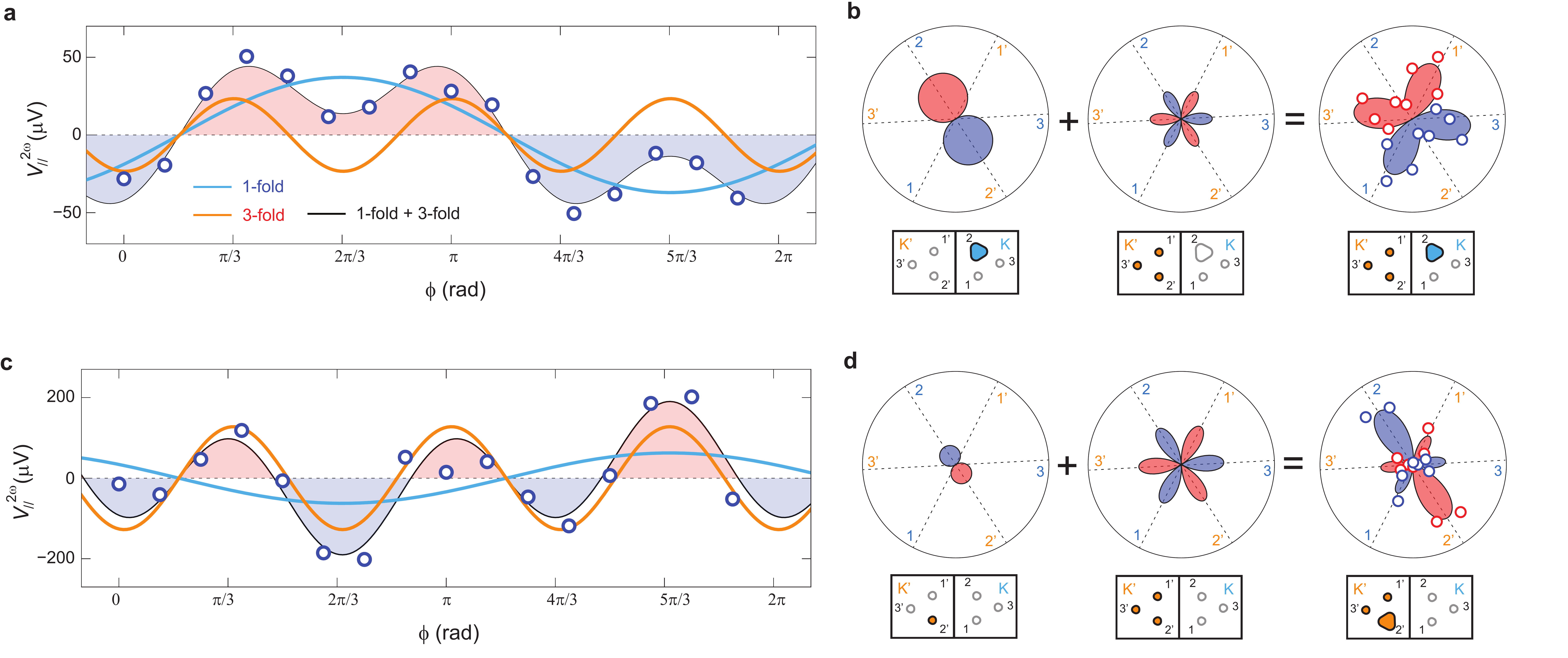}
\caption{\label{1and3fold}{\bf{The linear combination of one-fold and three-fold components.} }  (a-b) and (c-d) show two different types of mixed responses between one-fold and three-fold components.  (a) The angular dependence of the nonlinear transport response of the PVP/PMP phase. The best fit to the data (black solid line) is a linear combination of the one-fold (blue solid line) and three-fold (orange solid line) components. (b) The polar-coordinate plot of the one-fold and three-fold components as shown in panel (a) (left and middle panels). The rightmost panel shows the angular dependence of the actual data with the best fit.  In (a-b), the one-fold and three-fold components have opposite polarity. That is to say, the maximum positive response of the one-fold component occurs along the same azimuth angle as the maximum negative response of the three-fold component. In this case, we argue that the one-fold and three-fold components arise from the contribution of opposite valleys. The polar axis of the one-fold component is aligned along the occupied Fermi pocket $2$ in K valley, whereas the three-fold component arises from equally occupied pockets in K' valley. (c) The angular dependence of the nonlinear transport response, which is measured on the low-density side of transition boundary \textit{I}. This is the same angular dependence as shown in panel \textit{ii} of Fig.~\ref{fig2}b. The best fit to the data (black solid line) is a linear combination of the one-fold (blue solid line) and three-fold (orange solid line) components. (d) The polar-coordinate plot of the one-fold and three-fold components as shown in panel (c) (left and middle panels). The rightmost panel shows the angular dependence of the actual data with the best fit.  
In (c-d), the one-fold and three-fold components share the same polarity. This points towards a partially momentum-polarized state within the same valley.  While the three-fold component indicates charge carrier occupation amongst all three pockets in K' valley, a one-fold component with polar axis pointing along $2'$ suggests that extra charge carriers occupy Fermi pocket $2'$.
}
\end{figure*}

\subsection{Mixed angular dependence}

Fig.~\ref{1and3fold} shows two types of mixed angular dependence in the nonlinear transport response, which are best described by a linear combination between one-fold and three-fold components. We propose that the Fermi pocket occupation across two valleys can be accurately identified based on the polarity of the one-fold and three-fold components. 

In Fig.~\ref{1and3fold}a, the polar axis of the one-fold component is aligned along one of the maximum negative dips of the three-fold component. In this scenario, the one-fold (blue solid line) and three-fold (orange solid line) components having opposite polarities. This is a strong indication that the one-fold and three-fold components arise from carrier occupation in opposite valleys. Fig.~\ref{1and3fold}b shows the polar-coordinate plots of different angular components. The one-fold and three-fold component on the left hand side of the equal sign correspond to the blue and orange solid lines in Fig.~\ref{1and3fold}a, respectively. According to the polar plots, the one-fold component is consistent with a momentum-space instability in K valley, which induces all charge carriers to occupy pocket $2$. On the other hand, the maximum positive response of the three-fold component occurs along the azimuth directions of $1'$, $2'$, and $3'$, which points towards equal carrier occupation in valley K'. The carrier occupation of each angular component is demonstrated  in the schematic diagram in the bottom panels. As the angular components combined to fit the measured data, the carrier occupations combined to give rise to a partially valley- and momentum-polarized phase, PVP/PMP. 

In Fig.~\ref{1and3fold}c, the one-fold (blue solid line) and three-fold (orange solid line) components share the same polarity. The polar axis of the one-fold component is aligned along one of the maximum positive peak of the three-fold component. We argue that this angular dependence arise from partial pocket occupation from the same valley. Fig.~\ref{1and3fold}b shows the polar-coordinate plots of different angular components. According to the polar plots in Fig.~\ref{1and3fold}b, the one-fold component results from a momentum-space instability in K' valley where charge carriers flock to occupy pocket $2'$. At the same time, the maximum positive response of the three-fold component occurs along the azimuth directions of $1'$, $2'$, and $3'$, which suggests that all three pockets in valley K' are occupied. As such, the overall angular dependence of the nonlinear response gives rise to a valley-polarized state with partial momentum-polarization in K' valley, \textit{VP/PMP}.  

Notably, the \textit{VP/PMP} and PVP/PMP phases occurs on two sides of transition boundary \textit{I}. This confirms that transition \textit{I} separates different valley isospin orders.

\begin{figure*}[h]
\includegraphics[width=0.45\linewidth]{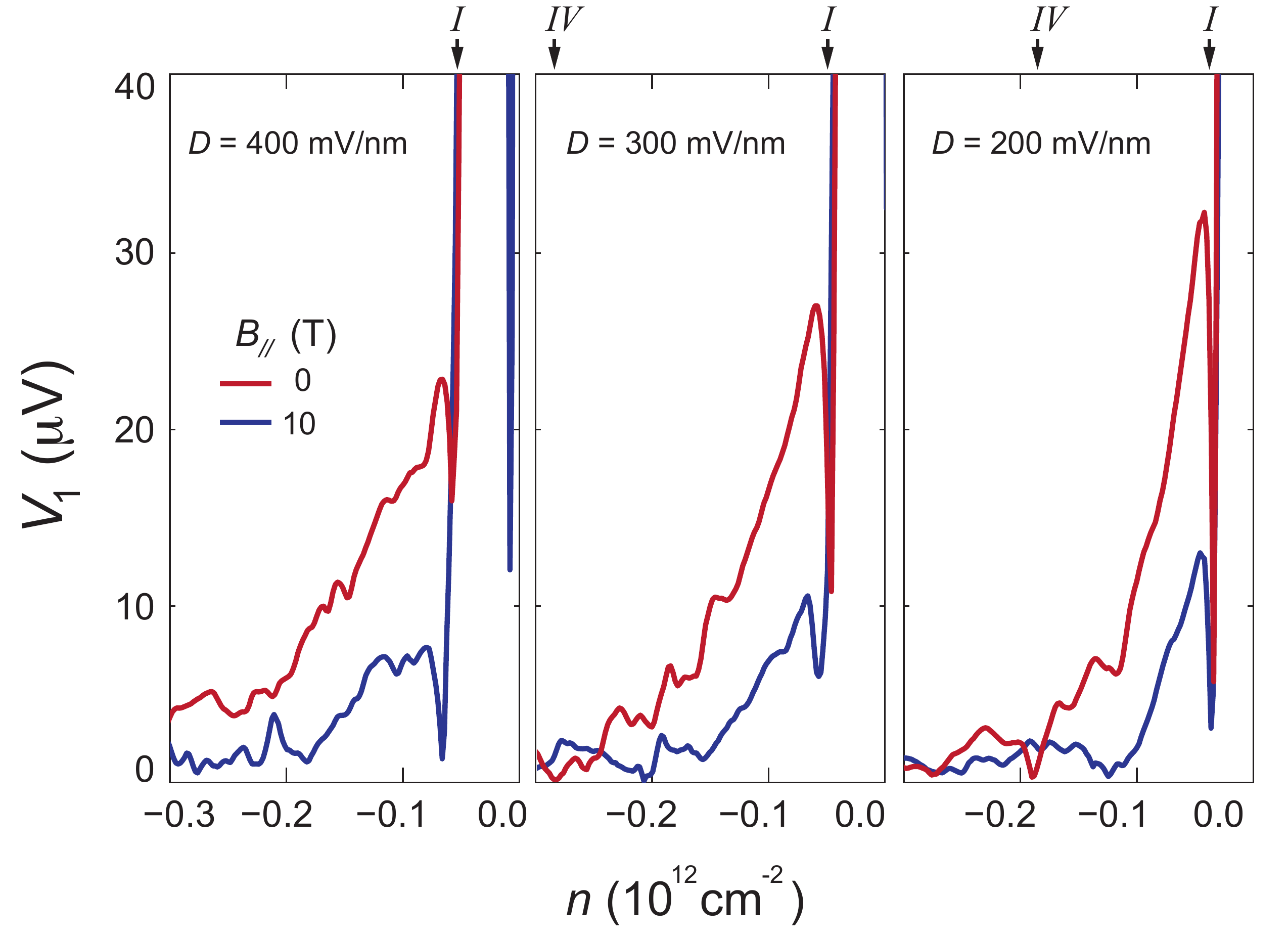}
\caption{\label{figBinplane}{\bf{In-plane $B$ dependence.} } $V_1$ as a function of charge carrier density measured at different \Bpara. The red (blue) solid line corresponds to \Bpara\ $= 0$ ($10$ T). Vertical arrows near the top axis mark the position of transition boundaries \textit{I} and \textit{IV}. In the density regime between these two boundaries, nonlinear response is unstable against \Bpara. On the other hand, nonlinear response on the low density side of \textit{I} remains robust up to \Bpara\ $=10$ T. This measurement further supports the influence of in-plane $B$ as shown in Fig.~\ref{fig4}e-g. }
\end{figure*}

\begin{figure*}[!b]
\includegraphics[width=0.79\linewidth]{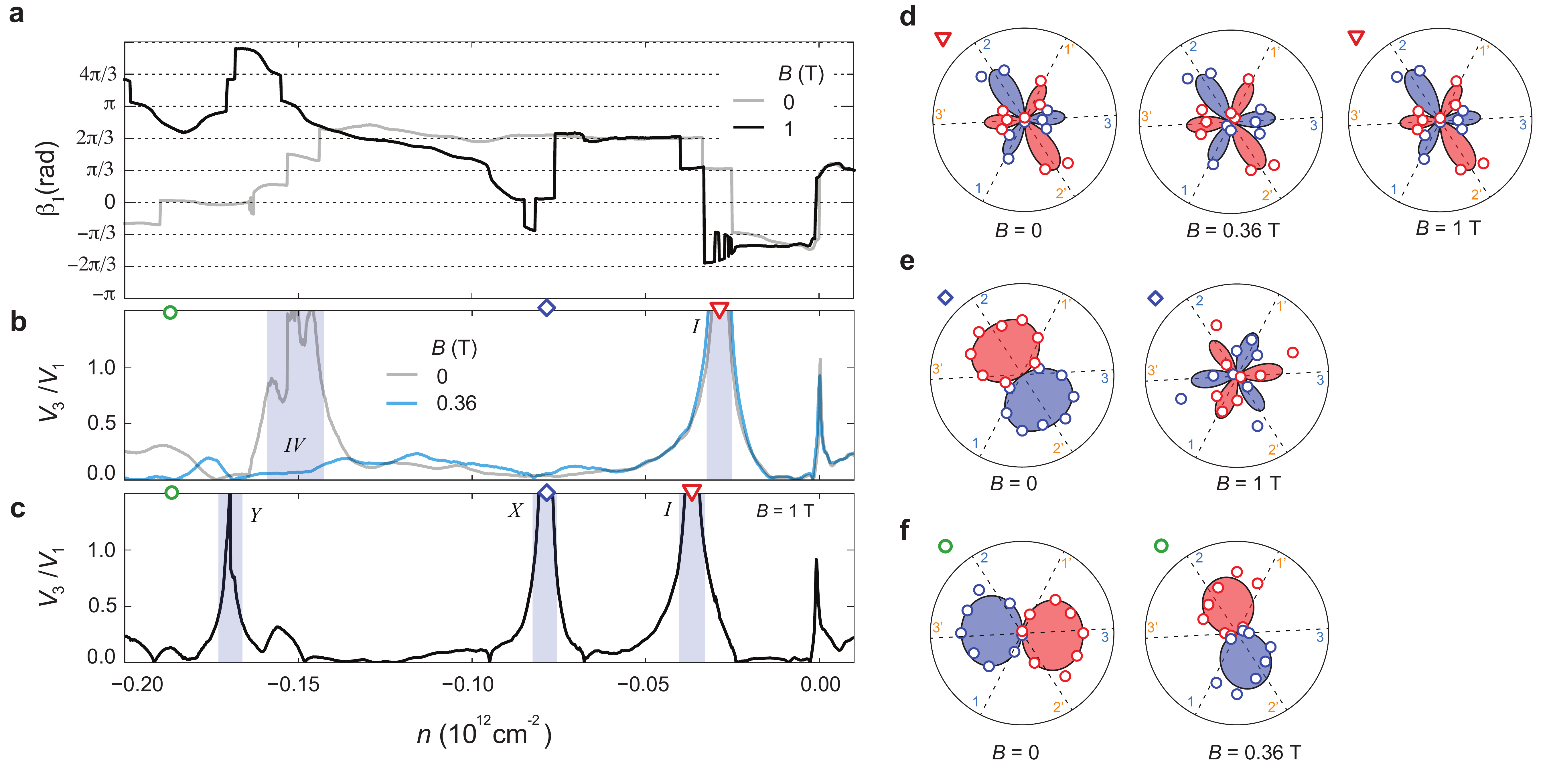}
\caption{\label{figBSI}{\bf{The influence of an out-of-plane $B$-field on valley and momentum polarization.} } The density dependence of (a)  $\beta_1$ measured at $B=0$ (black line) and $1$ T (grey solid line), (b) $V_3/V_1$ measured at $B=0$ (grey trace) and $B=0.36$ T (blue trace), (c) $V_3/V_1$ measured at $B=1$ T (black trace). Different transition boundaries show distinct $B$-dependence. At $B=1$ T, \textit{I} is pushed to slightly higher density, indicating that the MP$_1$ phase on the low-density side of \textit{I} becomes more stable compared to the PVP/PMP phases on the high-density side of \textit{I}. This is consistent with the fact that MP$_1$ is spin-polarized, whereas  PVP/PMP is spin-unpolarized. Moreover, transition  \textit{IV} is highly unstable against $B$. As shown in panel (b), transition \textit{IV} disappears completely at $B = 0.36$ T. At $B=1$ T, two new transitions appear near $n = -0.08 \times 10^{12}$ cm$^{-2}$ and $n = -0.17 \times 10^{12}$ cm$^{-2}$, evidenced by rotations in $\beta_1$ and peaks in $V_3/V_1$. These transitions are labeled \textit{X} and \textit{Y}, which arises from the coupling between momentum polarization and the $B$-field. (d) Polar-coordinate plots of nonlinear response measured at the MUP$_1$ phase near transition boundary \textit{I}. The density of the MUP$_1$ phase is marked by red triangles for different $B$. While the transition \textit{I} is pushed to a slightly higher density at \Bpara $=10$ T, the MUP$_1$ phase remains the same. (e) Polar-coordinate plots of nonlinear response measured at the PVP/PMP$_{12}$ phase, which is marked by blue diamonds.  At $B = 1$ T, the one-fold symmetric angular dependence of the PVP/PMP$_{12}$ phase evolves into a three-fold symmetric response. This is a strong indication that the PVP/PMP$_{12}$ phase is replaced by a momentum-unpolarized state.  (f) Polar-coordinate plots of nonlinear response measured on the high-density side of \textit{IV}. As transition \textit{IV} disappears at $B= 0.36$T, the polar axis shows a prominent rotation. All measurements are performed at $D = 150$ mV/nm.
 }
\end{figure*}

\subsection{The influence of an external magnetic field}

Both in-plane and out-of-plane magnetic field couples strongly to the momentum-space instability, as shown in Fig.~\ref{figB} and Fig.~\ref{fig4}. Here we provide more detailed discussion to these $B$-dependence.

Fig.~\ref{figBinplane} plots $V_1$ as a function of carrier density with (blue solid lines) and without (red solid lines) an in-plane magnetic field. The location of transition boundary \textit{I} and \textit{IV} are marked by vertical arrows near the top axis. The  PVP/PMP phases occupy the density range between \textit{I} and \textit{IV}. Across a wide range of $D$, the application of an in-plane magnetic field suppresses $V_1$ substantially in the PVP/PMP phases, whereas it has little effect on the MP$_1$ phase on the low density side of \textit{I}. The influence of an in-plane $B$ on the PVP/PMP phases offers further confirmation that the Zeeman-induced spin order competes against the instability in the valley and momentum channel. 

Similarly, an out-of-plane $B$-field couples strongly to the momentum-space instability. As shown in Fig.~\ref{figB}, a small $B$-field induces a cascade of transitions between different momentum-polarized state. While Fig.~\ref{figB} is measured from the high-density regime, where the Fermi surface is predominantly four-fold degenerate, the influence of an out-of-plane $B$ is  prominently observed in the low-density regime as well,  where the three-pocket model is applicable.  Fig.~\ref{figBSI} plots the density dependence of $\beta_1$ and $V_3/V_1$ across transition boundaries \textit{I} and \textit{IV} at different $B$-field.  According to Fig.~\ref{figBSI}a-c, transition \textit{I} shifts slightly towards higher density at $B=1$ T, which is evidenced by the peak position in $V_3/V_1$, as well as the jump in $\beta_1$. Independent of this shift, the three-fold symmetric angular dependence in the nonlinear transport response remains the same up to \Bperp $=1$T (Fig.~\ref{figBSI}d). On the other hand, transition \textit{IV} disappears completely at \Bperp $=0.36$ T. The stability of \textit{I} and \textit{IV} against \Bperp\ is similar compared to the temperature dependence in Fig.~\ref{figSIT}. With increasing temperature, \textit{I} remains mostly unchanged up to $T = 15$ K, whereas \textit{IV} is quickly pushed to higher density. The $B$ and $T$ dependence of transition boundary \textit{IV} remain an open question. 

At $B=1$ T, both $\beta_1$ and $V_3/V_1$ show two new transitions. In the density regimes of the PVP/PMP phases, the new transition labeled $X$ is highly consistent with a simultaneous transition in the valley and momentum channel. This transtion is accompanied by a density range with three-fold symmetric angular dependence in the nonlinear response, as shown in the right panel of Fig.~\ref{figBSI}e. At the same time, a jump in $\beta_1$ indicates a rotation in the polar axis across this transition. Similarly, another transition marked by $Y$ appears at $B=1$ T on the high-density side of \textit{IV}. $Y$ is distinct from \textit{IV}, since the latter is fully suppressed at $B=0.36$ T. Taken together, our observations suggest that an out-of-plane magnetic field has a prominent influence on the momentum-space instability, which impacts the Fermi surface contour. This adds a potential constraint for interpreting the Fermi surface contour based on magneto-oscillation alone.

\begin{figure*}[h]
\includegraphics[width=0.75\linewidth]{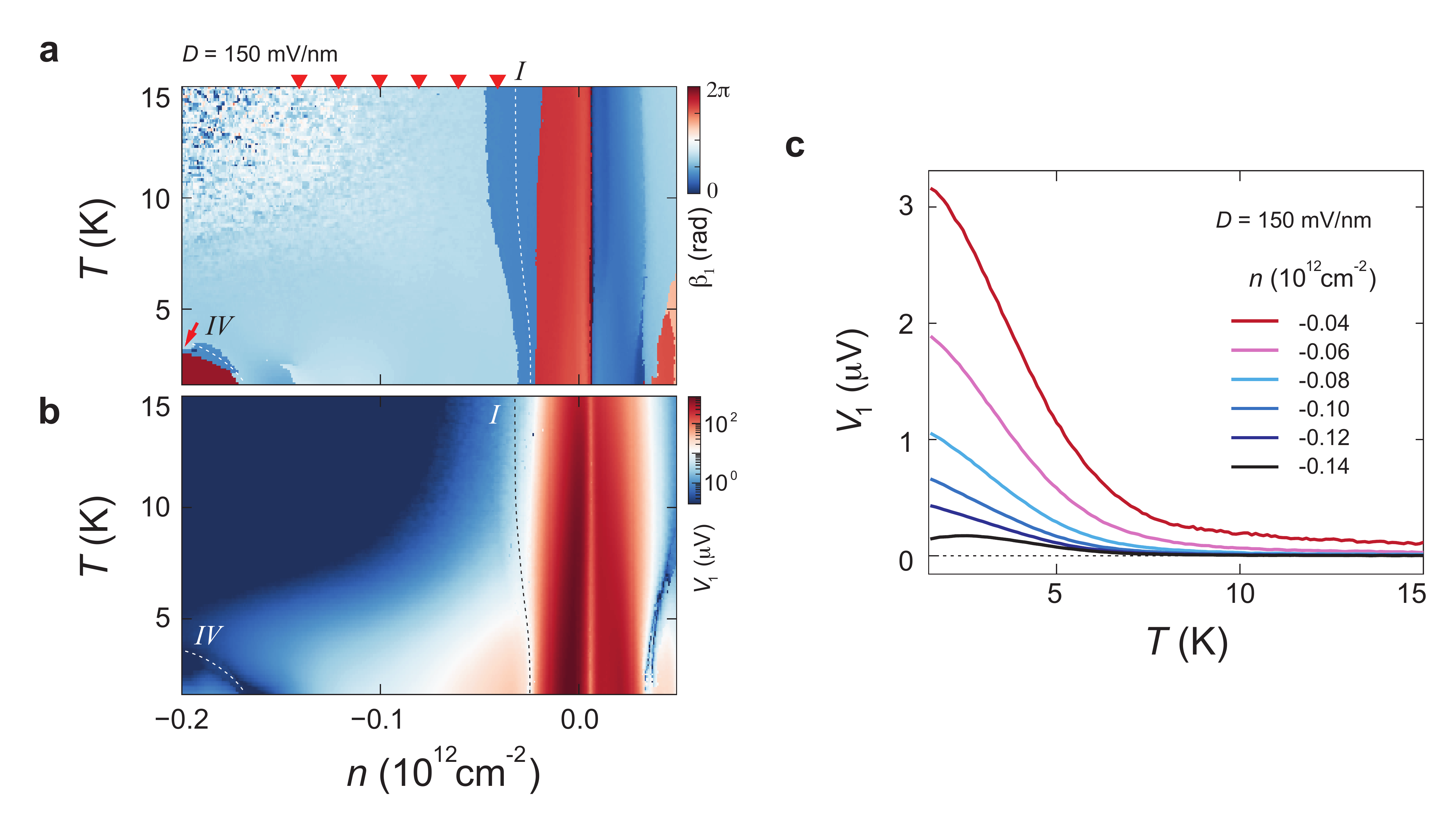}
\caption{\label{figSIT}{\bf{Temperature dependence at $B=0$.} } (a-b) $n-T$ map of (a) $\beta_1$ and (b) $V_1$ measured at $D = 150$ mV/nm. Transition boundaries \textit{I} and \textit{IV} are marked by sharp minima in $V_1$, which coincide with prominent rotation in the polar axis. These boundaries are marked with dashed lines in both panels (a) and (b).  (c) The temperature dependence of $V_1$ measured at different densiteis, which are marked by vertical red arrows near the top axis of panel (a). With increasing carrier density, $V_1$ onsets at a lower temperature, while exhibiting a smaller value in the low-temperature limit. This suggests that the strength of momentum polarization is linked to the Coulomb correlation strength.
 }
\end{figure*}

\subsection{The temperature dependence of momentum-space instability}

Fig.~\ref{figSIT}a-b plots $\beta_1$ and $V_1$ across the density-temperature map. $V_1$ offers a direct characterization for the strength of momentum polarization, whereas $\beta_1$ identifies transitions between different momentum-polarized states. The transition boundaries \textit{I} and \textit{IV} both correspond to sharp minima in $V_1$, which coincide with prominent rotation in the polar axis. \textit{IV} shifts to higher density with increasing $T$. While the width of transition \textit{I} becomes broader with increasing $T$, the location of \textit{I} is mostly independent of $T$. 

Away from the transition boundaries, the temperature dependence of momentum polarization exhibits a characteristic density dependence. Fig.~\ref{figSIT}c plots the temperature dependence of $V_1$ measured at different density. A much larger $V_1$ is observed at lower density, where the Fermi level is closer to the edge of the flat band, whereas $V_1$ diminishes with increasing $n$. The trend in density implies that the ground state is partially momentum polarized at higher density. The largest value of $V_1$ is observed in the MP phase on the low-density side of \textit{I}.

\subsection{Bernal BLG}

\noindent \textbf{$D$-induced Energy gap at the CNP:} 
Fig.~\ref{figD} shows transport characterization of BLG. At $D=0$, the sample is highly conductive near the CNP. This indicates that BLG is misaligned with hBN substrates. With increasing $D$, the emergence of an energy gap at the CNP is evidenced by the resistance peak. The width of this peak is roughly $\delta n = \pm 1 \times 10^{10}$ cm$^{-2}$, which is consistent with other BLG samples with hBN/graphite dual encapsulation ~\cite{Zibrov2017,Li.17b,Zhou2022BLG,Zhang2022BLG}.   This offers a strong indication of excellent sample quality.
\\

\begin{figure*}[h]
\includegraphics[width=0.45\linewidth]{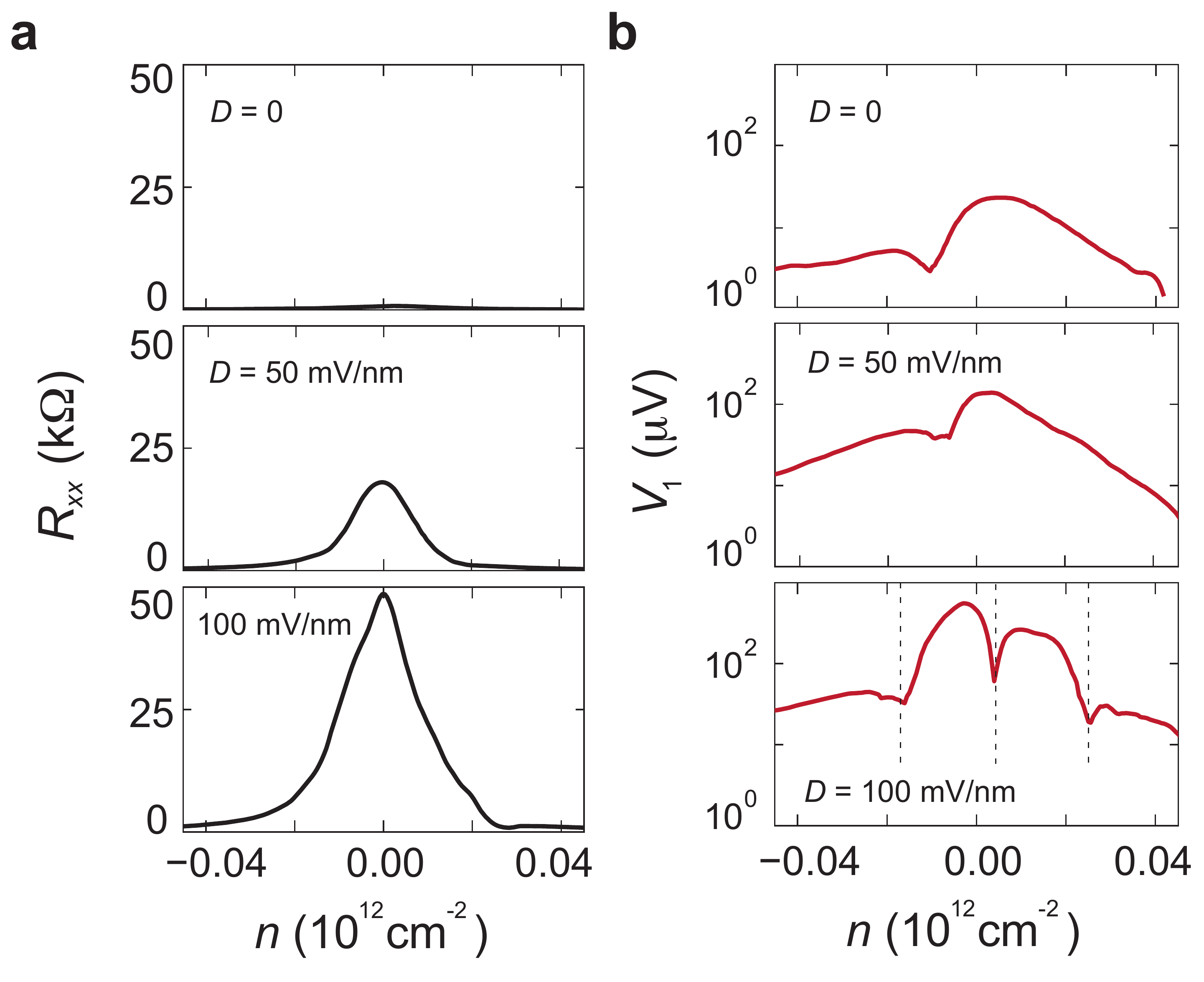}
\caption{\label{figD}{\bf{$D$-dependence in transport response.} } (a) Longitudinal resistance $R_{xx}$ and (b) nonlinear response \Vt\ as a function of density $n$ measured at different $D$-field.
 }
\end{figure*}

\noindent  \textbf{Valley and momentum polarization near $D=0$:}
As shown in Fig.~\ref{fig2}a, the angular behavior across the CNP depends on the value of $D$. At $D > 70$ mV/nm, the CNP coincides with simultaneous transitions in the valley isospin order and momentum polarization.  The sharpness of this transition suggests that the influence of the electron-hole charge puddle regime near the CNP has negligible impact on the momentum space instability. At $D < 70$ mV/nm,  the angular dependence of the nonlinear response remains mostly the same on both sides of the CNP (Fig.~\ref{fig2}a). The lack of rotation in the polar axis across the CNP could indicate a Coulomb-driven reconstruction in the energy band, which pushes the real charge neutrality point to around $n = -1 \times 10^{10}$ cm$^{-2}$. 

\vspace{0.3cm}
\noindent \textbf{Electron-hole symmetry:}
The angular dependence across the $n-D$ map exhibits excellent electron-hole symmetry. Across $A'$ and $B'$, the polar axis displays a $180^{\circ}$ rotation (Fig.~\ref{fig2}c), with the transition boundary showing a similar trajectory compared to its hole-doped counterpart. To account for the reversed charge carrier polarity on the electron-side of the phase space,  we define the polar axis to be aligned along maximum negative nonlinear response. For example, the one-fold symmetric angular dependence in regime $A'$ ($B'$) denotes carrier occupation of Fermi pocket $1$ ($1'$). As a result, the transition boundaries between $A$ and $A'$, $A'$ and $B'$ are associated with a reversal in both valley and momentum polarization. Both of these boundaries are accompanied by a small density regime of three-fold symmetric angular response (Fig.~\ref{fig2}a and e).

\begin{figure*}
\includegraphics[width=0.5\linewidth]{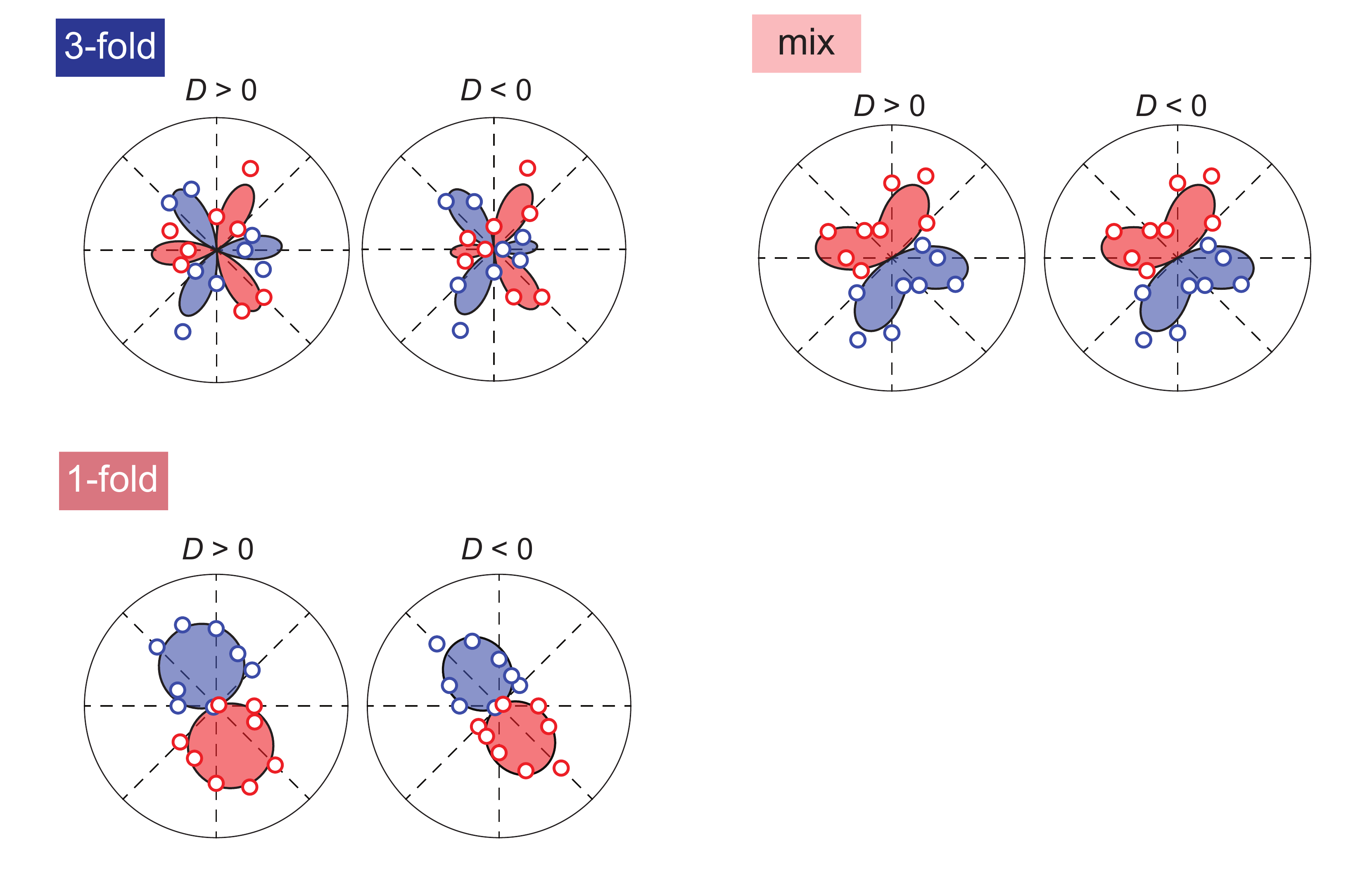}
\caption{\label{figDSI}{\bf{Reversing the electric field $D$.} } At the same carrier density, reversing the electric field $D$ does not change the angular dependence of the nonlinear response. This is a strong indication that sublattice polarization, which switches sign upon reversing $D$, is of secondary importance in stabilizing the one-fold and three-fold symmetric nonlinear response.
 }
\end{figure*}

\begin{figure*}[h]
\includegraphics[width=0.55\linewidth]{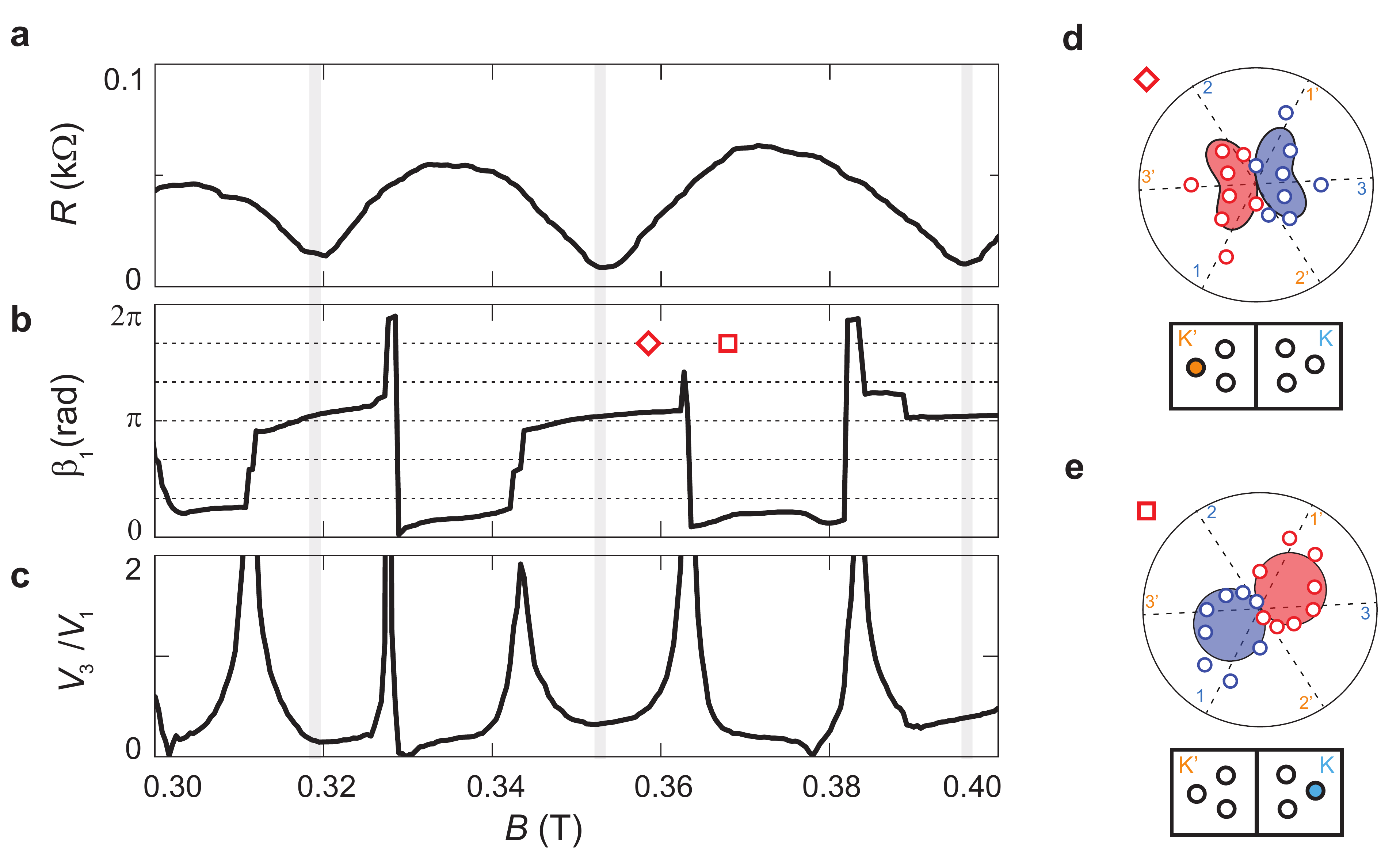}
\caption{\label{figB2fold} The $B$ dependence of (a) $R_{\parallel}$, (b) $\beta_1$, and (c) $V_3/V_1$ measured in the high-density regime of $n= -0.3 \times 10^{12}$ cm$^{-2}$ and $D=0$, which is the same density and electric field as Fig.~\ref{figB}. The $B$-dependence of $\beta_1$ and $V_3/V_1$ offers more insights into the $B$-induced transition marked by the vertical blue arrow in Fig.~\ref{figB}g. Vertical grey stripes mark the location of minima in \Rxx. According to panels (b) and (c), the rotation in the polar axis is accompanied by a sharp peak in $V_3/V_1$, indicating that the transition boundary coincides with a predominantly three-fold symmetric angular dependence in the nonlinear transport response. Furthermore, polar-coordinate plots in (d-e) show the angular dependence of the nonlinear response measured at the $B$-field value marked by red diamond and square, which are on two sides of a $B$-induced transition. The polar axis rotates by roughly $180^{\circ}$ across this transition. Combined, the angular dependence across the transition points towards a simultaneous transition in the valley- and momentum-polarization. 
 }
\end{figure*}

\newpage
\clearpage

\section{Materials and Method}

\subsection{Device Fabrication}

The Bernal BLG sample is doubly encapsulated with hexagonal boron nitride (hBN) and graphite crystals, following the same procedure as discussed in Ref.~\cite{Zibrov2017,Li.17b,Spanton2018}. All components of the structure are assembled from top to bottom using the same poly(bisphenol A carbonate) (PC)/polydimethylsiloxane (PDMS) stamp mounted on a glass slide. The sequence of stacking is: graphite as top gate electrode, $33$ nm thick hBN as top dielectric, graphite as contact, Bernal-stacked BLG, $57$ nm thick hBN as bottom dielectric, bottom graphite as bottom gate electrode. The entire structure is deposited onto a doped Si/SiO$_2$ substrate, as shown in Fig.~\ref{figDevice}a. A crystal of graphite directly contacts BLG, which is etched into eight individual contacts using standard nano-fabrication procedures, which includes electron-beam lithography and plasma etching with CHF$_3$/O$_2$. Each graphite contact is further connected with gold leads using electron beam deposition of the Cr/Au (2/100 nm) metal edge contacts.

\begin{figure*}[h]
\includegraphics[width=0.95\linewidth]{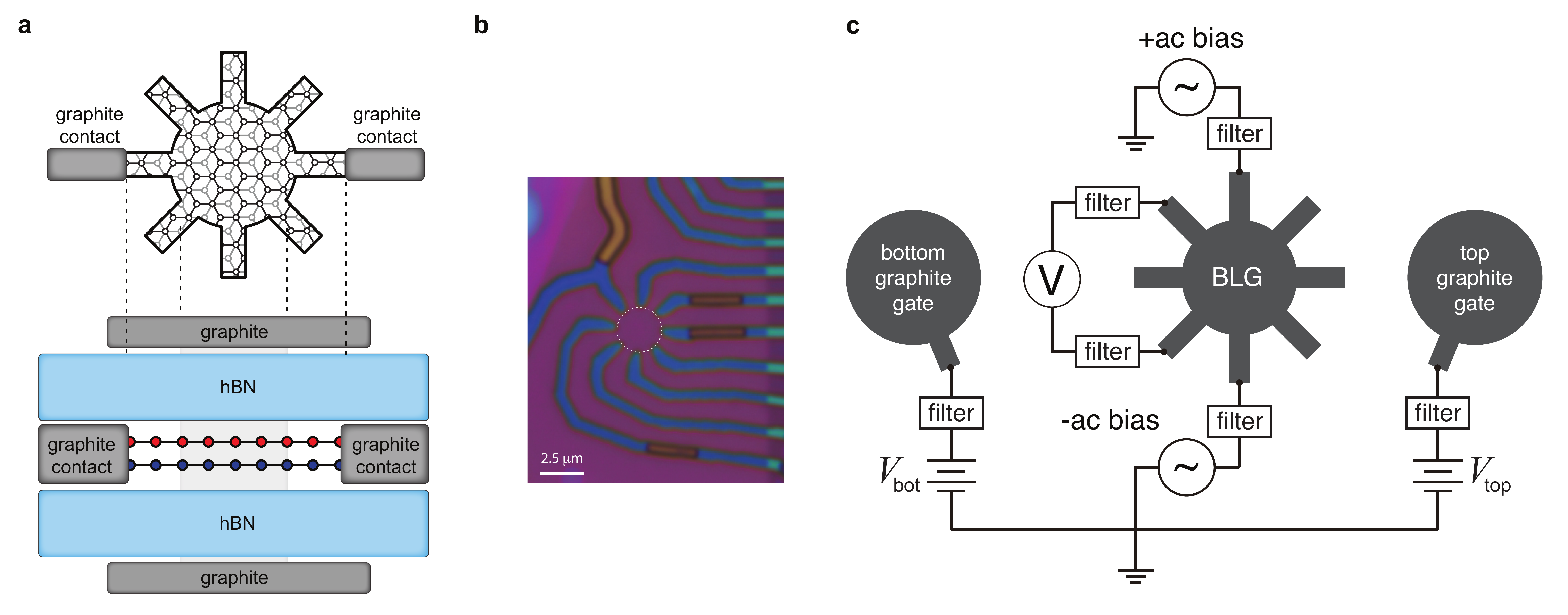}
\caption{\label{figDevice} {\bf{Sample geometry and angle-resolved transport measurement.} }   (a) Schematic top and side view the BLG sample patterned into the ``sunflower'' geometry. (b) Optical image of the BLG sample. The channel of the ``sunflower'' sample is highlighted by white dashed line.  (c) Schematic diagram of the  \arnt\ setup.  
Top and bottom gate electrodes, both made of graphite crystals,  cover the entire area of Bernal-stacked BLG, including parts of the electrical leads. This ensures that there is no junction within BLG. AC current bias is applied symmetrically to two sides of the BLG. This ensures that the center of the sample is placed at a virtual grounding potential.
}
\end{figure*} 

\subsection{$n-D$ phase space of BLG}

The dual-encapsulated geometry allows independent control of carrier density and electric field in Bernal BLG , $n$ and $D$. Such control is achieved by applying a DC gate voltage to top graphite electrode $V_{top}$, and bottom graphite electrode $V_{bot}$. $n$ and $D$ can be obtained using the following equations: 
\begin{eqnarray}
n_{BLG} &=& (C_{top}V_{top}+C_{bot}V_{bot})/e+n^0_{BLG}, \label{EqM1}\\
D_{BLG} &=& (C_{top}V_{top}-C_{bot}V_{bot})/2\epsilon_0, \label{EqM2}
\end{eqnarray} 
\noindent
where $C_{top}$ is the geometric capacitance between top graphite and BLG, $C_{bot}$ the geometric capacitance between bottom graphite and BLG. $n^0_{BLG}$ is the intrinsic doping of BLG. 

\subsection{Transport measurement}

The second-harmonic nonlinear transport response is measured by applying an AC current at a frequency of $13$ Hz.  The nonlinear response is measured at the second harmonic frequency $26$ Hz, between two contacts aligned parallel to the direction of the current flow using Stanford Research SR830 amplifier. The magnitude of the AC current is equal to or smaller than $60$ nA. To avoid the potential influence of contact resistance and thermal effect, the AC current is applied symmetrically across the sample, as shown in Fig.~\ref{figDevice}c. 

All measurements are performed in a BlueFors LD400 dilution refrigerator with a base temperature of $20$ mK. We have installed an external multi-stage low-pass filter on the mixing chamber, which is commercially available from QDevil. The filter contains two filter banks, one with RC circuits and one with LC circuits. The radio frequency low-pass filter bank (RF) attenuates above $80$ MHz, whereas the low frequency low-pass filter bank (RC) attenuates from $50$ kHz. The filter allows electrons to thoroughly thermalize with the mixing chamber, thus ensuring low electron temperature. 

%The angle dependence of the nonlinear transport measurement is obtained by flowing AC current along different azimuth angle. As discussed in Ref.~\cite{Zhang2022sunflower,Zhang2022valley}, a ``sunflower'' geometry with $8$ petals allows 16 azimuth angles of current flow, providing sufficiently high angle resolution to identify valley and momentum polarized states. 

\newpage

\end{widetext}
\end{document}